\documentclass[twocolumn]{autart}

\usepackage{graphicx}      
\usepackage{amssymb}  
\usepackage[shortlabels]{enumitem}  
\usepackage{array}
\usepackage{nicematrix}
\usepackage{color}
\usepackage{pifont}
\usepackage{tabularx} 
\usepackage{balance}
\usepackage{algorithm}

\usepackage{algorithmic}

\usepackage{hyperref}

\begin{document}
\begin{frontmatter}

\title{Nature-inspired dynamic control for pursuit-evasion of robots} 

\thanks[footnoteinfo]{This paper is supported by KTH Digital Futures.}

\author[First]{Panpan Zhou} \ead{panpanz@kth.se},
\author[Second]{Sirui Li} \ead{siruil@kth.se},
\author[Third]{Benyun Zhao} \ead{byzhao@mae.cuhk.edu.hk},
\author[Fourth]{Bo Wahlberg} \ead{bo@kth.se},
\author[First]{Xiaoming Hu*} \ead{hu@kth.se}


\address[First]{Department of Mathematics, KTH Royal Institute of Technology, Stockholm, Sweden}
\address[Second]{Department of Transport and System Analysis, KTH Royal Institute of Technology, Stockholm, Sweden }
\address[Third]{Department of Mechanical and Automation Engineering, The Chinese University of Hong Kong, Hong Kong}
\address[Fourth]{Division of Decision and Control Systems, KTH Royal Institute of Technology, Stockholm, Sweden}

\begin{abstract}                
The pursuit-evasion problem is widespread in nature, engineering, and societal applications. It is commonly observed in nature that predators often exhibit faster speeds than their prey but have less agile maneuverability. 
Over millions of years of evolution, animals have developed effective and efficient strategies for both pursuit and evasion. 
In this paper, we provide a dynamic framework for the pursuit-evasion problem of unicycle systems, drawing inspiration from nature. First, we address the scenario involving one pursuer and one evader by proposing an Alert-Turn control strategy, which consists of two efficient ingredients: a sudden turning maneuver and an alert condition for starting and maintaining the maneuver. We present and analyze the escape and capture results at two levels: a lower level of a single run and a higher level with respect to parameters' changes. In addition, we provide a theorem with sufficient conditions for capture.
The Alert-Turn strategy is then extended to more complex scenarios involving multiple pursuers and evaders by integrating aggregation control laws and a target-changing mechanism. By adjusting a ‘selfish parameter’, the aggregation control commands produce various escape patterns of evaders: cooperative mode, selfish mode, as well as their combinations. The influence of the selfish parameter is quantified, and the effects of the number of pursuers and the target-changing mechanism are explored from a statistical perspective. Our findings align closely with observations in nature. Finally, the proposed control strategies are validated through numerical simulations that replicate some chasing behaviors of animals in nature. 
\end{abstract}


\end{frontmatter}

\section{Introduction}
Pursuit-evasion is a ubiquitous phenomenon in nature with great appeal to researchers for a long history from many disciplines, including mathematics \cite{friedman2013differential}, engineering \cite{kolling2009pursuit,lozano2022visibility}, and biology \cite{weihs1984optimal,wilson2018biomechanics,wilson2013locomotion}. This problem has been extensively studied from various perspectives such as area defense \cite{walrand2011harbor,liang2019differential}, sports, and aircraft control \cite{turetsky2003missile,imado2005method}. 
The pursuit-evasion problem is also often formulated as a differential game in, for instance, \cite{chen2016multi,yan2022matching,zhou2023distributed,ho1965differential,lopez2019solutions,xu2022multiplayer,basimanebotlhe2014stochastic}, ranging from scenarios involving a single pursuer and evader to those with multiple pursuers and evaders. 
However, it is challenging to replicate the real-world behavior in pursuit-evasion, especially for the multi-agent setup and nonlinear systems. The game is theoretically difficult even with quadratic objective function.
Optimal pursuit-evasion strategies for the two-player game can also be obtained using geometric approaches \cite{getz1981two,chen2016multi,garcia2017geometric}. However, these strategies are generally difficult to extend to scenarios involving multiple players with nonlinear dynamics.

Different from existing studies on the pursuit-evasion problem, which approach it from the perspectives of game theory and motion geometry, this work examines the problem from a bio-perspective.
The pursuer and evader in nature correspond to the predator and prey, respectively, and in the engineering world, they can represent various robotics or agents. Typically, a pursuer runs faster than its targeted evader, for example, the cheetah, the fastest land animal, can catch impala with ease. However, a faster pursuer usually has less agile maneuvers, meaning it has a larger turning radius as it turns at a higher speed. In contrast, the evader has more agile maneuvers \cite{wilson2013locomotion}. This agility gives evaders a critical advantage, allowing them to avoid capture. In nature, prey rely on these agile maneuvers as a key survival strategy.

The investigations and studies on the \textit{maneuverability} have been in discussion for several decades. The relative importance of speed and maneuverability in predator-prey chases was investigated in \cite{howland1974optimal}, \cite{webb1983speed}.  Using numerical methods, this work determined the values of relative radius and velocity that allow prey to escape. However, two significant assumptions were made: both the predator and prey initially move along the same straight line, and the turning radius is independent of velocity. The turning radius is even assumed to be constant in \cite{howland1974optimal}.

There are some aspects to reconsider such a problem. (1) Model precision: Linear models \cite{li2023predator,li2016dynamics}, which generally describe agents as mass points, are insufficient for accurately describing the dynamics of animals and robots which are inherently rigid body systems. A more precise model is needed to account for natural limitations such as velocity and turning radius in both pursuers and evaders. (2) Relaxing unrealistic assumptions: Removing the unrealistic assumptions made in prior studies \cite{howland1974optimal,weihs1984optimal,bopardikar2009cooperative,oyler2016dominance,patsko2004families} presents a significant challenge. These assumptions simplify the problem but fail to capture realistic scenarios.
(3) Impact of maneuverability: It is vital to demonstrate the maneuverability of pursuers and evaders. Even though the pursuer benefits from higher speed and the evader has quicker-turning maneuverability, results of capture and escape depend on their respective settings. The question of how different settings of the pursuer and evader influence  the result is still unanswered. (4) Group dynamics and target switching: In nature, pursuit-evasion often involves groups of predators and prey. For instance, wildebeests tend to escape in aggregation when chased by wild dogs, and predators may switch targets if their current one is too difficult to catch. Extending strategies to systems with multiple pursuers and evaders is therefore necessary. Moreover, it is still not clear how the number of pursuers and the target-changing mechanism will affect the result.  

In this paper, 
we aim to develop a dynamic escape and pursuit framework that broadly approximates general chasing and escape phenomena in nature from a bio-inspired perspective, instead of formulating the problem as a differential game, and to disclose some clues of natural laws. The main contributions are as follows.

We propose a dynamic framework for robots to model the live-or-death pursuit-evasion strategies of animals. To this end, we model the pursuers and evaders as unicycle systems, characterized by linear velocity and angular velocity as control inputs, instead of linear systems in \cite{li2023predator,li2016dynamics}.
Firstly, we introduce an escape strategy, termed the Alert-Turn algorithm, which consists of two simple but efficient phases: (i) the suddenly turning maneuver, and (ii) the alert condition for starting and maintaining the maneuver. No matter in which phase, robots adhere to constraints on velocities and the anti-correlation between angular velocity and linear velocity \cite{wilson2013locomotion}. Then, we extend the Alert-Turn algorithm to address the more complex scenario of multiple pursuers and evaders, that are not considered in \cite{li2023predator,li2016dynamics,howland1974optimal,weihs1984optimal,bopardikar2009cooperative,patsko2004families,webb1983speed,weihs1984optimal}. Motivated by natural phenomenons that preys escaping in cooperation and selfish modes, or their combinations \cite{lee2006dynamics,li2024intelligent}, and that predators change their targets now and then \cite{jackson2006evolution}, our design incorporates similar strategies to replicate these phenomena. Although the work \cite{li2016dynamics} proposed a similar Alert-Turn algorithm, it relied on linear system models and only discussed the case of one-pursuer-one-evader. From the control perspective, the robots have two discrete control modes, with transition determined by the alert distance. From the mathematical perspective, the modeling and algorithm is hybrid and highly nonlinear, thus it is highly challenging to derive analytical solutions. To solve this problem, we analyze it and present results from a statistical perspective.

Then, from the proposed framework, we have the following dynamic properties of the systems. For the Alert-Turn algorithm, (1) we provide some escape and capture examples at the level of individual run, and the reasons for successful capture and escape are explained. (2) At a higher level of results with respect to parameters' changes, we investigate the effects of key parameters, including the centripetal acceleration, tangential acceleration, initial conditions, and the alert distance. Our findings suggest that the evader should initiate turning maneuvers as early as possible within the alert distance and escape with the maximum acceleration. For pursuit-evasion with multiple robots, (1) we illustrate the evaders' diverse escape patterns of cooperative type and selfish type, and the pursuers' capture pattern of target-changing. (2) At a higher level, we quantify the dispersion degree with respect to a selfish parameter, and present how the number of pursuers and target-changing mechanism influence the capture outcomes. It shows that capture time decreases as the number of pursuers increases. For optimal performance, the pursuers should frequently detect targets, but avoid excessively frequent or infrequent target changes.

\begin{table*}[h]
	\begin{center}
		\caption{Notations of the model and the Alert-Turn algorithm}
	\begin{tabular}{cc}
		\hline\hline
		\textbf{Notations}    & \textbf{Physical Meaning} \\
        $x, y, \theta$ & $x$ position, $y$ position, and the orientation of the player\\	
        $v$ & linear velocity of the player\\
        $w$ & angular velocity of the player\\	
		$V_p^{\max}$, $V_e^{\max}$ &  the maximum linear velocity of the pursuer and the evade, respectively  \\
		$W_p^{\max}$, $W_e^{\max}$ &  the maximum angular velocity of the pursuer and the evader, respectively  \\
		$r_p$, $r_e$ &  the maximum centripetal acceleration of the pursuer and the evader, respectively\\
		$a_p$, $a_e$     &  the maximum tangential acceleration of the pursuer and the evader, respectively \\
		$c_p$      &  $c_pV_p^{\max}$ is the minimum velocity of the pursuer when decelerating \\
		$c_e$      &  $c_eV_e^{\max}$ is the minimum velocity of the evader when decelerating\\
		$\varepsilon_1$ & the alert distance for the evader  to take and  maintain the agile maneuvers\\
		$\varepsilon_2$ & the capture radius of the pursuer \\
		$\Delta d_{pe}$ & equals to $\sqrt{(x_p-x_e)^2+(y_p-y_e)^2}$, the relative distance between the pursuer and evader \\
		$\Delta \theta_{pe}$ & equals to ${\rm atan}\frac{y_e-y_p}{x_e-x_p}$, the relative angle between the pursuer and the evader\\
		$t_f$ & the terminal time of simulation\\
		\hline\hline
	\end{tabular}	
\label{notations}
	\end{center}
\end{table*}

Last but not least, to verify the effectiveness of our strategies, we compare our results with some chasing behaviors of animals in nature, from the perspective of time and distance. It shows that our strategy perfectly replicates their behaviors.


The rest of the paper is organized as follows: Section~\ref{sec2} describes the problem. Section~\ref{sec3} presents the Alert-Turn algorithm, and analyzes the results of a single run and higher level of results 
with respect to parameters' changes. Section~\ref{sec4} gives the strategies for pursuit-evasion problem involving multiple robots. Replication of animals' chasing behaviours is given in Section~\ref{sec5}. Some conclusion remarks and future works are given in Section~\ref{sec6}.

\section{Problem Description}  \label{sec2}
The objective of this article is to provide effective pursuit and escape strategies that approximate general pursuit-evasion behaviours from a nature-inspired perspective. In predator-prey chasing scenarios frequently observed in nature, the pursuer typically runs directly toward the evader as fast as possible, and the evader moves straight away from the pursuer. Once the distance between the pursuer and the evader shrinks to a certain threshold (denoted as $\varepsilon_1>0$), the evader suddenly turns left or right to evade capture. 

Given the challenges of capturing and recording sufficient hunting behaviors in nature, we model the dynamics and design strategies of both the pursuer and evader from a bio-inspired perspective. It allows us to uncover some natural laws, which not only provide insights into more complex animal behaviors but are also critical for practical implementations in robotics.
We are interested in answering the following questions: How can we effectively model the behaviors of predators and prey, particularly the quick-turning maneuvers observed during hunts? What strategies can an evader employ to outmaneuver a faster pursuer? What are the primary factors that influence whether an evader successfully escapes, and how do these factors operate? 

\section{Formulation and Design of Control laws}\label{sec3}
Before introducing the strategies, we summarize the main notations used throughout this article along with their physical meanings in Table \ref{notations}.

\subsection{Models of the Pursuer and the Evader}
We use the well-known unicycle model to represent the dynamics of both the pursuer and the evader for several compelling reasons. First, the turning maneuvers inherent in pursuit-evasion scenarios naturally result in circular motion, where the turning radius is positively correlated with velocity. Second, during hunting or evasive maneuvers, animals predominantly move in the direction of their heading. Therefore, the unicycle model is particularly suitable for accurately describing and mimicking these movements, as it allows control over both linear and angular velocity, closely reflecting the dynamics of real-world animal behaviors.

Specifically, we describe the pursuer and the evader as
\begin{equation} \label{dynamics}
	\begin{aligned}
		\dot x_p&= v_pcos(\theta_p),~~~ \dot y_p= v_psin(\theta_p),~~~ \dot {\theta}_p = w_p, \\
		\dot x_e&= v_ecos(\theta_e),~~~ \dot y_e= v_esin(\theta_e),~~~ \dot {\theta}_e = w_e,
	\end{aligned}
\end{equation}
where the subscripts $p$ and $e$ denote the pursuer and evader, respectively. ${\rm col}(x,y):=[x~ y]^{\rm T}\in \mathbb{R}^2$ represents the position of players' center of mass, $\theta$ is the orientation, and $u={\rm col}(v,w)\in \mathbb{R}^2$ is the control input with $v$ as the linear velocity and $w$ as the angular velocity. 

Different from \cite{li2016dynamics},  this work considers velocity constraints for the players. Specifically, let $V_p^{\max}$ and $V_e^{\max}$ denote the maximum linear velocity, and $W_p^{\max}$ and $W_e^{\max}$ denote the maximum angular velocity. It is straightforward to have the constraints
\begin{equation}
	\begin{aligned}
		0\le v_p(t)\le V_p^{\max},&~~~ 0\le v_e(t)\le V_e^{\max}, 	\\
		w_p(t)\le W_p^{\max},& ~~~w_e(t)\le W_e^{\max}.
	\end{aligned}
\end{equation}
Negative values for the linear velocity are not allowed, as the pursuer and evader move only in the direction of their respective headings.

Typically, a predator runs faster than its prey, but a faster predator usually has less agile maneuver, resulting in a larger turning radius. This leads to the constraints
\begin{equation}
	V_p^{\max}> V_e^{\max}, ~~~~~ W_p^{\max}< W_e^{\max}. \label{con_V}
\end{equation}

\subsection{The Control Strategies of the Pursuer and the Evader} \label{sec3.2}
In nature, the escape strategy of the evader usually consists of two efficient phases, i.e., the sudden turning maneuver, and an early alert mechanism for starting and maintaining the maneuver. Thus, the strategy we design is denoted as the Alert-Turn algorithm. Initially, both the pursuer and the evader move straightly, while once their distance shrinks to the \textit{alert-distance} $\varepsilon_1$, that is, $\Delta d_{pe}\le \varepsilon_1$, the evader takes an agile turning maneuver to escape, and the pursuer subsequently takes the same maneuver. This mechanism thus separates the escape strategies into two phases: long-distance phase and short-distance phase. The algorithm is briefly presented in Algorithm~\ref{tab:ATA}, and its details are introduced in the following sections.

\begin{algorithm}
    \caption{Alert-Turn Algorithm for the pursuit-evasion with one pursuer and one evader }
\begin{algorithmic}
    \IF{$\Delta d_{pe}(t)\le \varepsilon_2$} 
        \STATE $v_p=0$, $v_e=0$, $w_p=0$, $w_e=0$
        \RETURN
    \ELSIF{ $\Delta d_{pe}(t)>\varepsilon_1$ }
        \STATE the evader and pursuer take strategies (\ref{evd_long})--(\ref{pur_long})
    \ELSE
        \STATE the evader and pursuer take strategies (\ref{v-short})--(\ref{w-short})
    \ENDIF
\end{algorithmic}\label{tab:ATA}
\end{algorithm}

As described bellow, 
the entire pursuit-evasion process involves multiple acceleration and deceleration stages. We denote by $\bar t_p$ and $\bar t_e$ the time instants at which acceleration and deceleration events are triggered for the pursuer and evader, respectively. It is obvious that $\bar t_p$ and $\bar t_e$ will be dynamically updated throughout the process as new events occur.

When the evader is in \textbf{long-distance phase} ($\Delta d_{pe}(t)>\varepsilon_1$), it attempts to escape as fast as possible in the direction opposite to the line connecting its position with the pursuer's. Hence, the evader's long distance strategy is modeled as
\begin{equation} 
	\begin{aligned}
		v_e(t) &= {\rm sat}(v_e(\bar t_e)+a_e (t-\bar t_e),V_e^{\max}), \\
		w_e(t) &= -{\rm sgn}\big(\theta_e(t)-\Delta \theta_{pe}(t),\min (r_ev_e^{-1},W_e^{\max})\big),
	\end{aligned} \label{evd_long}
\end{equation}
where ${\rm sat}(\cdot)$ is the saturation function defined as
\begin{equation*}
	{\rm sat}(x,y)=\left\{\begin{aligned}
		&0, ~~~~~{\rm if} ~x\leq 0 \\
		&x, ~~~~~{\rm if} ~0<x\leq y \\
		&y, ~~~~~{\rm otherwise.}
	\end{aligned}
	\right.
\end{equation*}
Note that the linear velocity is not allowed to be negative, as the player is restricted to forward motion. The ${\rm sgn}(\cdot)$ function is defined as 
\begin{equation*}
	{\rm sgn}(x,\sigma)=\left\{   \begin{aligned}
		&\sigma\cdot {\rm sign}(x),~~~~~~~{\rm if}~ \|x\|>\sigma^{1/\gamma} \\
		& {\rm sign}(x) \cdot \|x\|^{\gamma}, ~~{\rm otherwise}
	\end{aligned}
	\right.
\end{equation*}
where $0<\gamma<1$ is a constant. Note that ${\rm sgn}(x,\sigma)\le \sigma$. It can be proved that $w_e(t)$ converges to the desired value $\theta_e(t)-\Delta \theta_{pe}(t)$ in a finite time under the ${\rm sgn}(\cdot)$ function \cite{fu2018global}.
It is also worth noting that the angular velocity in (\ref{evd_long}) is constraint not only by the maximum angular velocity $W_e^{\max}$ but also by the linear velocity and the centripetal acceleration \cite{wilson2013locomotion}, as higher velocities usually lead to weak agile maneuver ability.
Since $\Delta \theta_{pe}$ denotes the orientation of the vector pointing from the pursuer to the evader, the angular velocity designed in (\ref{evd_long}) drives the evader to escape along the line connecting the pursuer and the evader. 
For simplicity and to avoid ambiguity, we omit the explicit dependence on time $t$ in the subsequent discussion.

The pursuer's larger turning radius, caused by its higher linear velocity during a chase, can be disadvantageous for capture. In nature, predators often decelerate to adjust angles until they faces or nearly face their prey. 
Inspired by this behavior, we use the same idea to mitigate the effects of a large turning radius. The long-distance strategy of the pursuer is thus designed as
\begin{equation}\label{pur_long}
	\begin{aligned}
		v_p&= \left\{ 
		\begin{aligned}
			&{\rm sat}(v_p(\bar t_p)+a_p(t\!-\!\bar t_p), V_p^{\max}),~{\rm if}~\|\theta_p-\Delta \theta_{pe}\| \!\leq\! \bar \theta \\
			&{\rm sat2}(v_p(\bar t_p)-a_p(t-\bar t_p, c_pV_p^{\max}),~~ {\rm otherwise}  
		\end{aligned}
		\right.\\
		w_p &= -{\rm sgn}\big(\theta_p-\Delta \theta_{pe},\min (r_p v_p^{-1},W_p^{\max})\big),
	\end{aligned}
\end{equation}
where $c_p\in [0,1)$ is a constant.
Here, the pursuer decelerates to adjust its orientation until it almost faces the evader (i.e., $\|\theta_p-\Delta \theta_{pe}\|\leq \bar \theta$). Then, it accelerates to chase the evader. The ${\rm sat2}(\cdot)$ is defined as
{\begin{equation*}
		{\rm sat2}(x,y)=\left\{\begin{aligned}
		&y, ~~~~~{\rm if} ~x\le y \\
		&x, ~~~~~{\rm otherwise.}
	\end{aligned}
	\right.
\end{equation*}
In other words, $c_pV_p^{\max}\le v_p\le V_p^{\max}$ under the ${\rm sat2}(\cdot)$ function, where $c_pV_p^{\max}$ denotes the minimum linear velocity during decelerating. 
The intuition behind the designs of (\ref{evd_long}) and (\ref{pur_long}) is that the pursuer chases the evader along a straight line, with both moving at their respective maximum velocities \cite{howland1974optimal}.

In the \textbf{short-distance phase} ($\varepsilon_2<\Delta d_{pe}(t)\le \varepsilon_1$), the evader takes the sharp-turn strategy, and so is the pursuer. As commonly observed in nature, predators decelerate when turning, which would enable much tighter turns (like what cheetahs do in hunting \cite{wilson2013locomotion}).
Therefore, both the pursuer and the evader decelerate first to decrease their turning radius.
Once noticing the sharp turn of the evader, the pursuer will take the same strategy to follow the evader. So, the pursuer and evader decelerate as
\begin{equation}\label{v-short}
	\begin{aligned}
		v_e &= \left\{ 
		\begin{aligned}
			&{\rm sat}\big(\max\{0,v_e(\bar t_e)-a_e(t-\bar t_e)\}, V_e^{\max}\big), \\& \qquad \qquad \qquad \qquad \qquad \qquad \qquad {\rm if}~ v_e> c_e V_e^{\max} \\
			& {\rm sat}(v_e(\bar t_e)+a_e(t-\bar t_e), V_e^{\max}),~~~~~~~~~~{\rm otherwise},
		\end{aligned}
		\right.  \\
		v_p &= \left\{ 
		\begin{aligned}
			&{\rm sat}\big(\max\{0,v_p(\bar t_p)-a_p(t-\bar t_p)\}, V_p^{\max}), \\& \qquad \qquad \qquad \qquad \qquad \qquad \qquad {\rm if}~ v_p> c_p V_p^{\max}\\
			& {\rm sat}(v_p(\bar t_p)+a_p(t-\bar t_p), V_p^{\max}),~~~~~~~~~~{\rm otherwise}
		\end{aligned}
		\right.
	\end{aligned}
\end{equation}
where $c_e\in [0,1)$ and $c_p\in [0,1)$ are constants with $c_e V_e^{\max}$ and $c_p V_p^{\max}$ denoting the minimum velocities during deceleration. Under the linear velocity command (\ref{v-short}), both players decelerate initially to enable sharp turning and then accelerate to speed their escape or chasing.
The angular velocities of players are designed as
\begin{equation} \label{w-short}
	\begin{aligned}
		w_e & = \left\{ 
		\begin{aligned}
			&{\rm sat}(-{\rm acot}(k_e \Delta d_{pe})v_e^{-1},\min \{r_ev_e^{-1},W_e^{\max}\}), \\& \qquad \qquad \qquad \qquad \qquad \qquad {\rm if}~\theta_e-\theta_p>0 \\
			&{\rm sat}({\rm acot}(k_e \Delta d_{pe})v_e^{-1},\min \{r_ev_e^{-1},W_e^{\max}\}), \\& \qquad \qquad \qquad \qquad \qquad \qquad  \quad  {\rm otherwise}
		\end{aligned}
		\right.  \\
        w_p&={\rm sign}(w_e){\rm sat}({\rm acot}(k_p \Delta d_{pe})v_p^{-1},\min \{r_p v_p^{-1},W_p^{\max}\})\\
	\end{aligned}
\end{equation}
where $k_e>0$ and $k_p>0$ are constants, ${\rm sign}(w_e)$ is the sign function with ${\rm sign}(w_e)=1$ if $w_e>0$, it equals to -1 if $w_e<0$, otherwise, it is 0. 
The angular velocity commands in (\ref{w-short}) enable the evader to turn left or right based on the relative orientation between the two players. To successfully capture the evader, the pursuer has the same turning direction as the evader. 
It is also worth noting that the real-time angular velocity is negatively correlated with the distance between the pursuer and the evader. The physical meaning behind it is that as the distance decreases, the evader turns more sharply to evade capture.  The function ${\rm acot}(\cdot)$ can also be replaced by other decreasing functions with positive values, such as $e^{-k\cdot \Delta d_{pe}}$.

\subsection{Results with Respect to a Single Run}

\begin{figure}	
	\centering
	\includegraphics[width=0.415\textwidth]{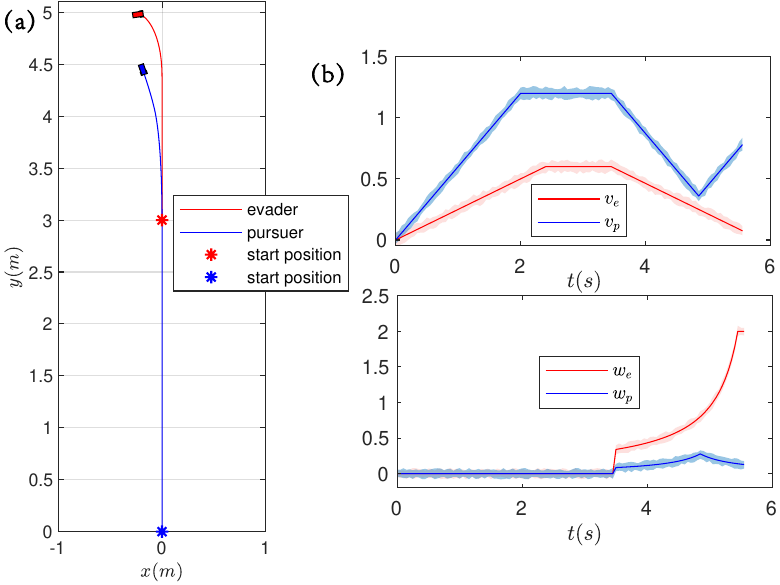}
	\includegraphics[width=0.445\textwidth]{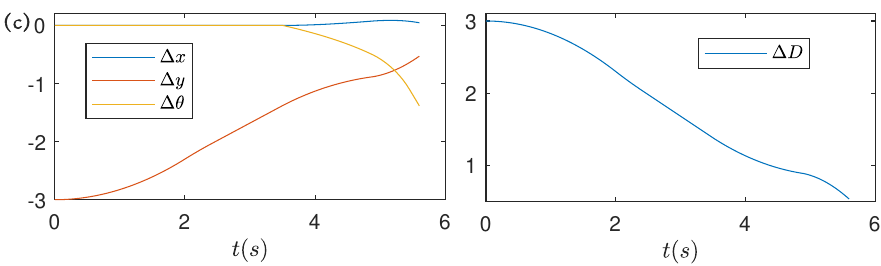}
	\caption{Illustration of the Alert-Turn Algorithm. $\varepsilon_1=1.4$, $\varepsilon_2=0.04$, $V_e^{\max}=0.6$, $V_p^{\max}=1.2$, $W_e^{\max}=2$, $W_p^{\max}=1$, $a_e=0.25$, $a_p=0.6$, $r_e=0.2$, $r_p=0.1$, $t_f=5.8$.}
	\label{ATA}
\end{figure}

\begin{figure*}
	\centering
	\includegraphics[width=0.47\textwidth]{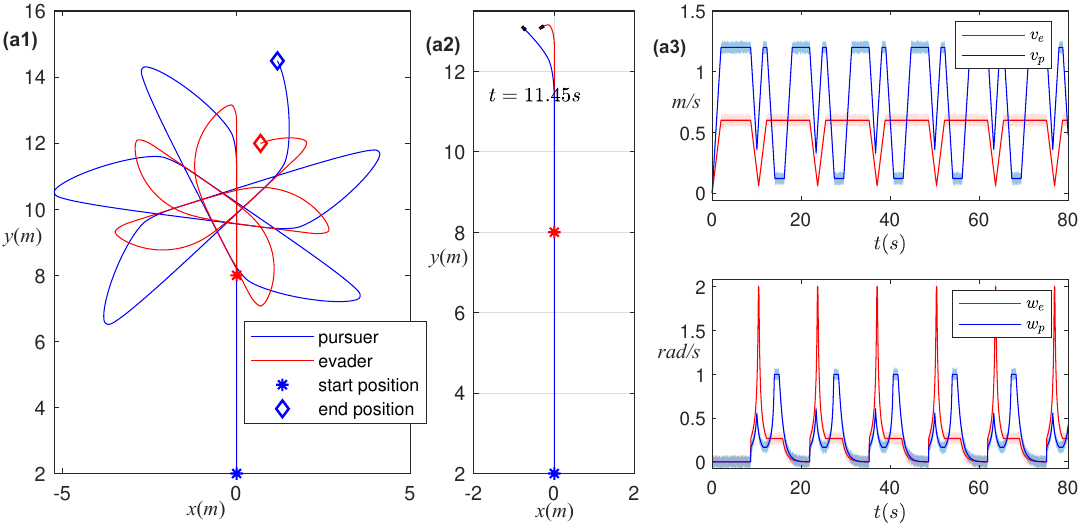}~~~~~
	\includegraphics[width=0.47\textwidth]{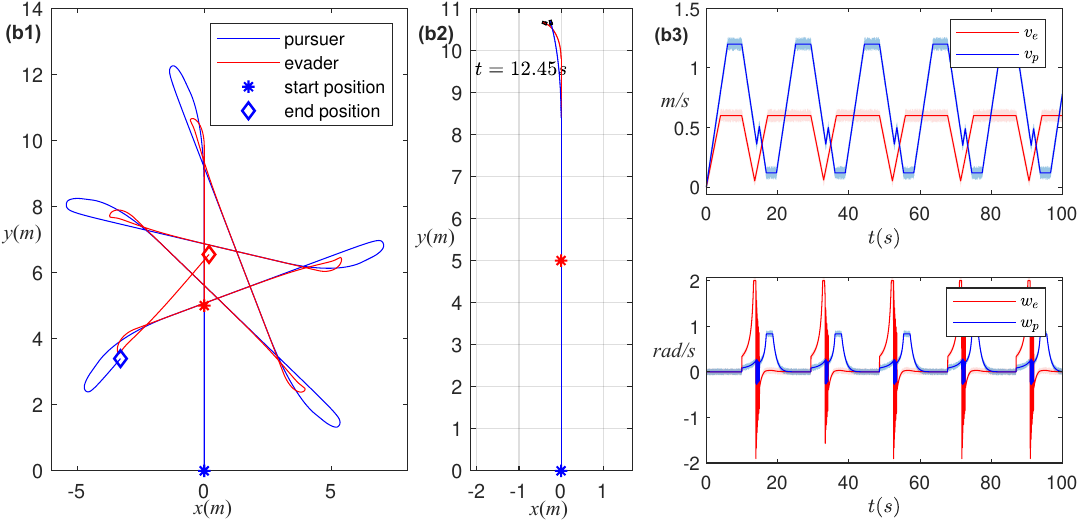}
	\includegraphics[width=0.47\textwidth]{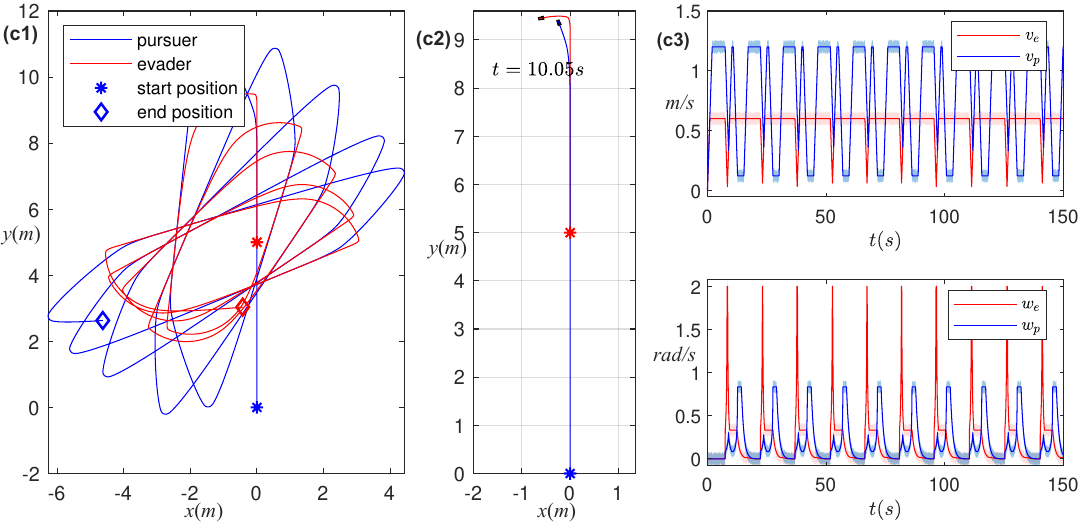}~~~~~
	\includegraphics[width=0.47\textwidth]{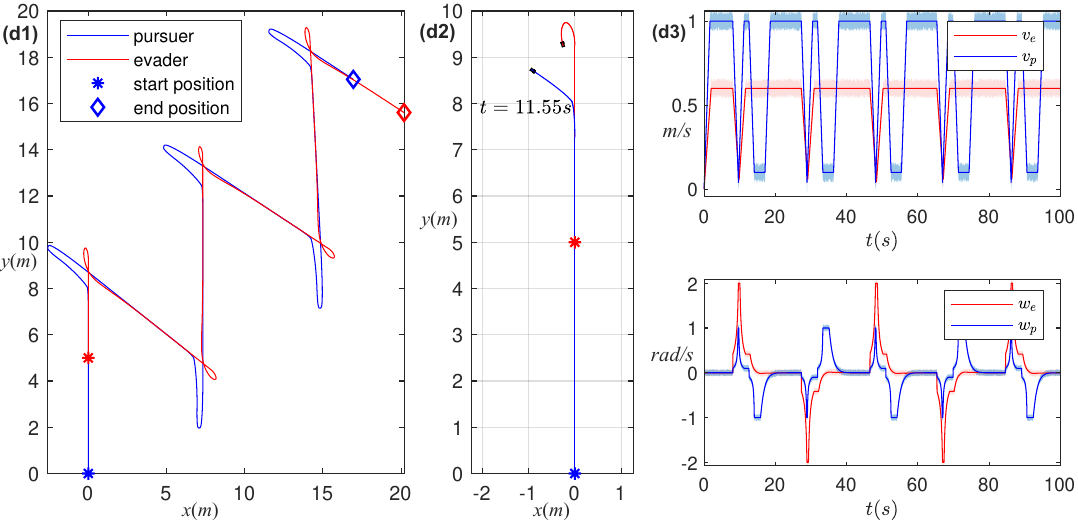}
	\caption{Illustration of escape patterns. The first column figures show the trajectories of agents. The second column shows the moment the evader escapes at its first turning maneuver. The third column figures present how the velocities evolve with time. (a) $\varepsilon_1=1.4$, $\varepsilon_2=0.04$, $r_e=0.16$, $r_p=0.2$, $c_e=0.1$, $c_p=0.3$, $a_e=0.3$, $a_p=0.6$, $t_f=80$. 
    (b)  $\varepsilon_1=1.4$, $\varepsilon_2=0.04$, $r_e=0.2$, $r_p=0.1$, $c_e=0.1$, $c_p=0.3$, $a_e=0.15$, $a_p=0.2$, $t_f=100$. 
    (c) $\varepsilon_1=1.4$, $\varepsilon_2=0.04$, $r_e=0.2$, $r_p=0.1$, $c_e=0.1$, $c_p=0.3$, $a_e=0.6$, $a_p=0.6$, $t_f=150$. 
    (d)  $\varepsilon_1=2$, $\varepsilon_2=0.05$, $r_e=0.25$, $r_p=0.1$, $c_e=0.05$, $c_p=0.3$, $a_e=0.3$, $a_p=0.6$, $t_f=100$.}
	\label{escape}
\end{figure*}

\begin{figure*}
	\centering
	\includegraphics[width=0.46\textwidth]{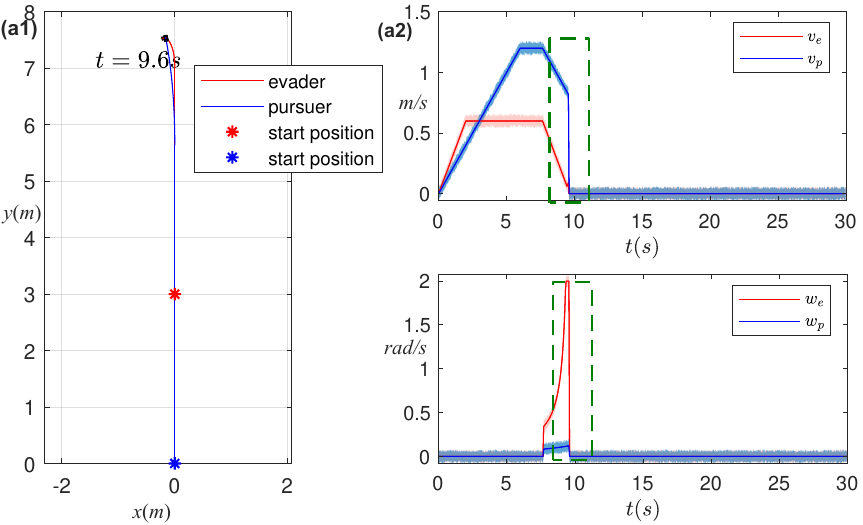}~~~~~
	\includegraphics[width=0.46\textwidth]{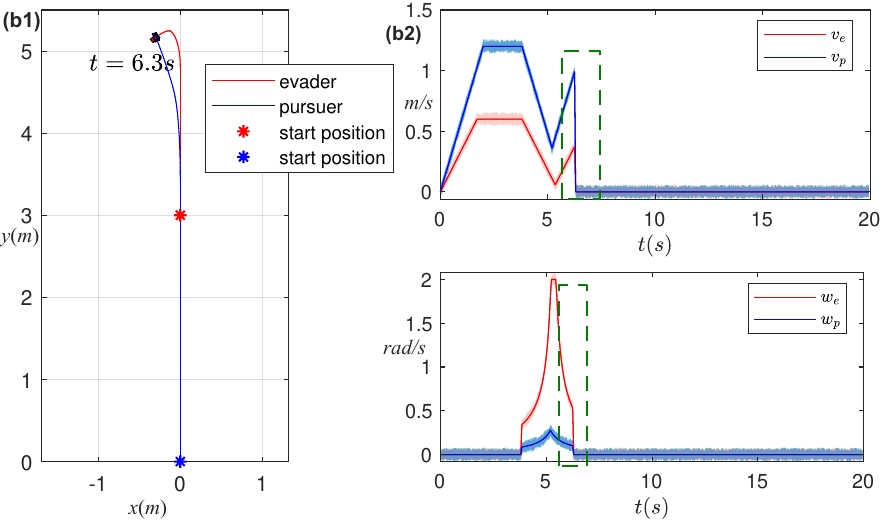}
	\includegraphics[width=0.46\textwidth]{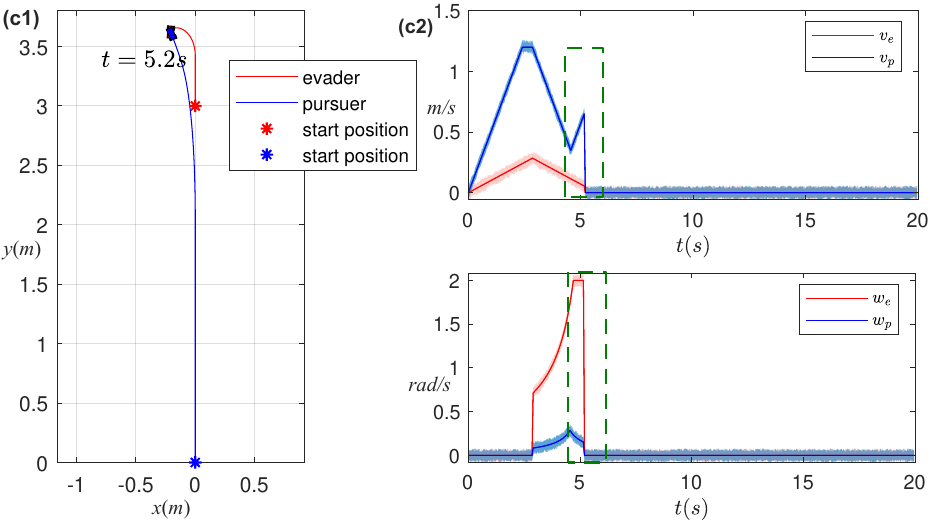}~~~~~
	\includegraphics[width=0.46\textwidth]{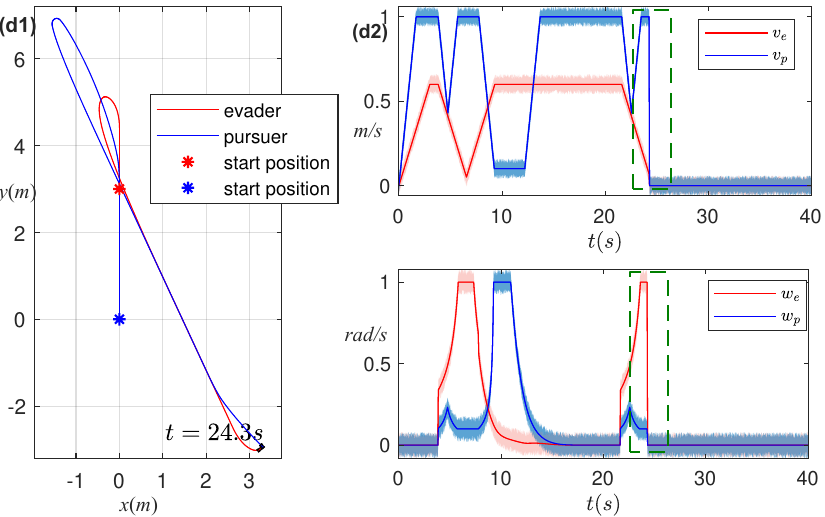}
	\caption{Illustration of capture patterns. The first column figures show the trajectories of agents. The second column presents how the velocities evolve with time. (a)  $\varepsilon_1=1.4$, $\varepsilon_2=0.04$, $r_e=0.23$, $r_p=0.1$, $c_e=0.1$, $c_p=0.05$, $a_e=0.3$, $a_p=0.2$, $t_f=30$. 
    (b) $\varepsilon_1=1.4$, $\varepsilon_2=0.04$, $r_e=0.2$, $r_p=0.1$, $c_e=0.1$, $c_p=0.3$, $a_e=0.35$, $a_p=0.6$, $t_f=20$. 
    (c)  $\varepsilon_1=1.4$, $\varepsilon_2=0.04$, $r_e=0.2$, $r_p=0.1$, $c_e=0.1$, $c_p=0.3$, $a_e=0.1$, $a_p=0.5$, $t_f=20$. 
    (d)  $\varepsilon_1=1.4$, $\varepsilon_2=0.1$, $r_e=0.2$, $r_p=0.1$, $c_e=0.1$, $c_p=0.43$, $a_e=0.2$, $a_p=0.6$, $t_f=40$.}
	\label{capture}
\end{figure*}

Denote $t_d$ as the time when the evader is captured ($t_d= \infty$ if the pursuer can never capture the evader). For easy illustration of the simulations, if the pursuer cannot capture the evader until the end of the designated deadline $t_f$, we assign the value of $t_d$ as $t_f$, i.e., $t_d=t_f$. 

Fig.~\ref{ATA} illustrates an example of the Alert-Turn algorithm, where a faster pursuer chases a vulnerable evader. At the beginning, the evader accelerates to move straight away from the pursuer, but they are getting closer because the pursuer moves faster. When the distance decreases to the alert distance ($\varepsilon_1=1.4$) at about $t=3.45s$, the evader begins to adopt the turning maneuver to avoid capture. During this maneuver, the evader has a sharp decrease in linear velocity and a corresponding sharp increase in angular velocity. The pursuer has similar dynamics as the evader, however, it is notable that it has lower angular velocity than the evader. These dynamics closely resemble the observed hunting behaviors of predators and prey in nature \cite{wilson2013locomotion}. 
In order to better illustrate our results, we provide the videos for simulations of this article in https://youtu.be/m2AR\_iiRku8. 

In what follows, some examples of escape patterns and capture patterns are given and analyzed. In these examples, the maximum linear velocity of the pursuer ($V_p^{\max}=1.2$) is twice that of the evader ($V_e^{\max}=0.6$), while the maximum angular velocity of the evader ($W_e^{\max}=2$) is twice that of the pursuer ($W_p^{\max}=1$).

Fig. ~\ref{escape} shows some escape patterns under different settings, which evolve to be periodic. Although the linear velocity of the pursuer is twice that of the evader, the evader still survives owing to the strategy. In example (a), the pursuer misses the evader because the evader has larger centripetal acceleration (so is deceleration) which leads to quicker turning at the beginning of maneuver (see Fig.~(a2) for reference). The evader survives in (d1) for the same reason. However, compared with (a), the evader in (d1) has larger centripetal acceleration and a lower $c_e$, allowing it to turn more sharply during the maneuver (see Fig.~(d2)). 
In (b), the evader turns faster than the pursuer but less faster than the evaders in (a) and (d). 
In (c), the evader benefits from a larger tangential acceleration, enabling it to decelerate and turn more effectively (see Fig.~(c2) for reference).

Some capture patterns are presented in Fig.~\ref{capture}. In example (a), even though the evader takes the turning maneuver, it is happen to be captured as it decelerates (see the area denoted by the green box in (a2)). This occurs because the pursuer also decelerates and executes a turning maneuver effectively. In (b), although the evader turns faster at the beginning (it has higher deceleration), it is captured after completing its deceleration phase and beginning to accelerate (see (b2)). In example (c), the pursuer has larger tangential acceleration. At the moment of capture, the pursuer has completed its deceleration phase and started accelerating, while the evader is still decelerating to take turns. In example (d), the evader survives at the first maneuver but is eventually captured as the pursuer adjusts its angle. This example shows the complexity and high non-linearity of the Alert-Turn algorithm. It is noted that the pursuer has higher velocity than the evader at the time of capture. The result aligns with the findings of cheetahs that they run slightly faster than their prey during the manoeuvering phase to capture agile and quick-turning targets \cite{wilson2013locomotion}.

Next we provide sufficient conditions for successful capture. 
\newtheorem{theorem}{Theorem}

\begin{theorem} \label{them1}
    For a given capture radius $\varepsilon_2$, assume $V_p>V_e$ and $W_e^{\max}>W_p^{\max}$.
    Let $\Delta d_{pe}(t_0)=d_0\le \varepsilon_1$, $\theta_e(t_0)=\pi/2$, $\theta_p(red{t_0})=\theta_p\in [\pi/2-\gamma,\pi/2+\gamma]$. If 
     \begin{enumerate}[topsep=-5pt]
     \item[i)] $r_p$ and $r_e$ satisfy $r_p\ge V_pW_p^{\max}$, $r_e\ge V_eW_e^{\max}$; 
    \item[ii)] $k_p$ and $k_e$ satisfy ${\rm acot}(k_p \varepsilon_1)v_p^{-1}\ge W_p^{\max}$, \\${\rm acot}(k_e\varepsilon_1)v_e^{-1}\ge W_e^{\max}$; 
    \item[iii)] $V_p-a_p\frac{\pi}{2 W_p^{\max}}\ge c_pV_p^{\max}$, $V_e-a_e\frac{\pi}{2W_p^{\max}}\ge c_e V_e^{\max}$; 
    \item[iv)] $W_p^{\max}< W_e^{\max} \le 3W_p^{\max}$; 
    \item[v)] $\varepsilon_1 < \varrho_1 + \sqrt{\varepsilon_2^2 -\varrho_2^2}$, where $\varrho_1 = V_p-V_e-2\frac{a_e}{(W_e^{\max})^2}-\frac{a_e\pi}{2W_p^{\max}W_e^{\max}}+(1-\frac{\pi}{2})\frac{a_p}{(W_p^{\max})^2}$, $\varrho_2 = 2\frac{V_e}{W_e^{\max}}-\frac{V_p}{W_p^{\max}} + \frac{a_p}{(W_p^{\max})^2} - \frac{a_e}{(W_e^{\max})^2}$,     
    \end{enumerate}
    then the evader will be captured at time $T$ with $T\in (0,\frac{\pi}{2W_p^{\max}}]$ for small enough $\gamma$.
\end{theorem}

\newtheorem{remark}{Remark}
\begin{remark}
The proof is given in the appendix.
The conditions (i) and (ii) guarantee that the evader and pursuer turn with their maximum angular velocities. Under condition (iii), both players decelerate as $t\in (0,\frac{\pi}{2W_p^{\max}}]$. Conditions (iv)-(v) ensure that the relative distance between the pursuer and the evader decreases over $t\in (0,\frac{\pi}{2W_p^{\max}}]$ and it is less than $\varepsilon_2$ at $t=\frac{\pi}{2W_p^{\max}}$. 
By choosing $a_p$ and $a_e$ small enough, (v) is easily satisfied.
We note that $\varrho_1>0$ since $V_p>V_e$, and $\varrho_2$ can be small if $V_e>V_p/2$. 
An illustrative example is given as follows: $\varepsilon_2=1$, $d_0=\varepsilon_1=1.31$, $V_p=2$, $V_e=1.6$, $W_p^{\max}=1$, $W_e^{\max}=2$, $\theta_p=\theta_e=\pi/2$, $r_p=3$, $r_e=4$,  $k_p$, $k_e$, $a_p$ and $a_e$ are small enough. 
    
\end{remark}

\subsection{Results with Respect to the Parameters' Ranges} \label{sec3.4}

The outcome is actually a function of the parameters in Table~\ref{notations}.
Figs.~\ref{param123}--\ref{param7} illustrate the pursuit-evasion results across different parameter ranges, which disclose some clues of natural laws and provide strategic guidance for both pursuers and evaders to enhance their likelihood of capture or escape.

\begin{figure}
	\centering
	\includegraphics[width=0.48\textwidth]{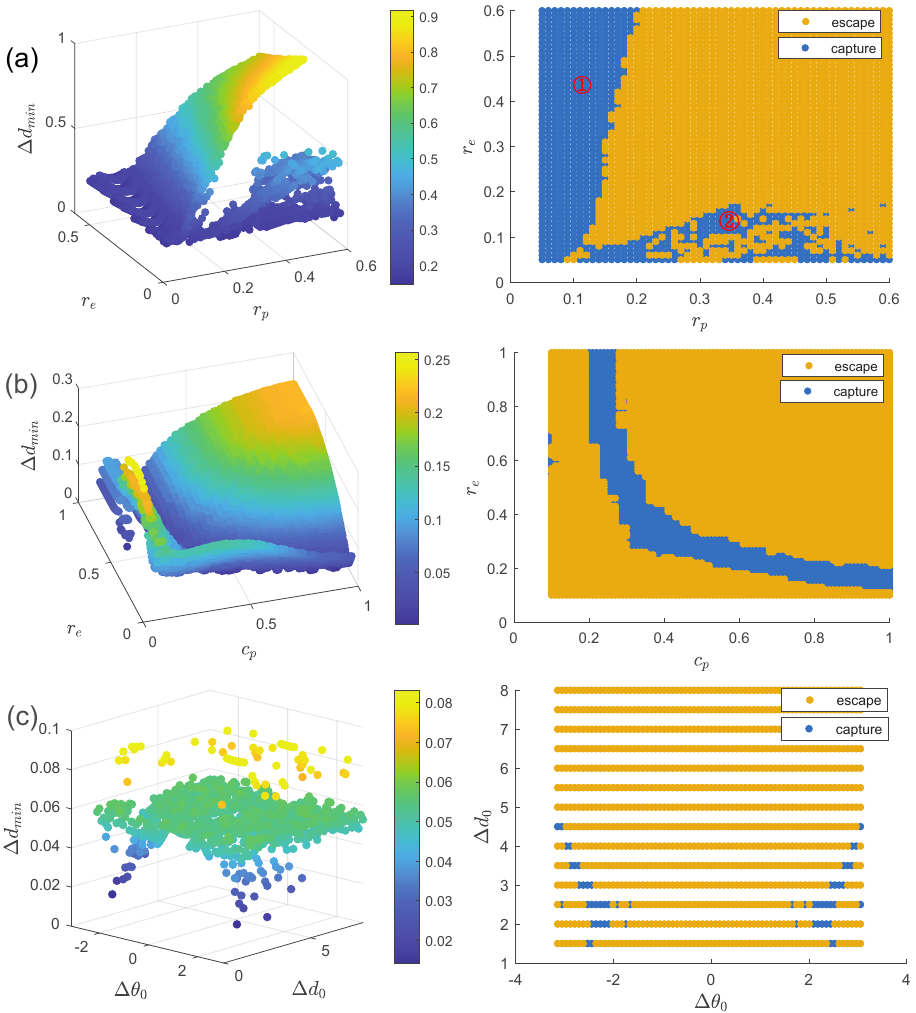}
	\caption{Simulation results with respect to the parameters' ranges. The common parameter values are $V_e^{\max}=0.6$, $V_p^{\max}=1.2$, $W_e^{\max}=2$, $W_p^{\max}=1$, $\varepsilon_1=1.4$, $\varepsilon_2=0.04$, $a_e=0.3$, $a_p=0.6$, $t_f=120$. Left: The minimum distance during the pursuit-evasion process with respect to different parameters. Right: The corresponding escape and capture results. (a) Results with respect to the ranges of $r_e$ and $r_p$ given $c_e=0.1$, $c_p=0.3$. (b) Results with respect to the ranges of $r_e$ and $c_p$, given $r_p=0.1$, $c_e=0.1$. (c) Results with respect to the ranges of initial conditions $\Delta \theta_0$ and $\Delta d_0$, given $c_e=0.1$, $c_p=0.3$, $r_e=0.2$, $r_p=0.1$.}
	\label{param123}
\end{figure}

\begin{figure}
	\centering
	\includegraphics[width=0.48\textwidth]{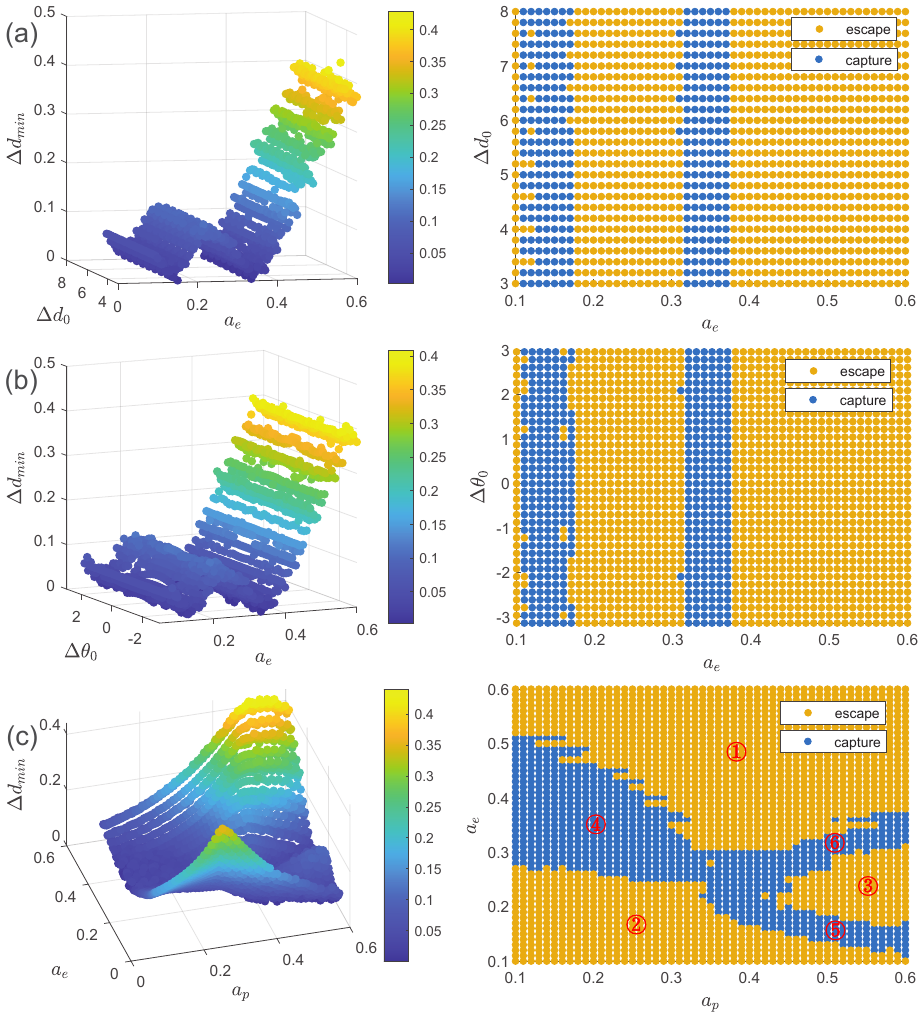}
	\caption{Simulation results with respect to the parameters' ranges. The common parameter values are $V_e^{\max}=0.6$, $V_p^{\max}=1.2$, $W_e^{\max}=2$, $W_p^{\max}=1$, $\varepsilon_1=1.4$, $\varepsilon_2=0.04$, $r_e=0.2$, $r_p=0.1$, $c_e=0.1$, $c_p=0.3$, $t_f=120$. Left: The minimum distance during the pursuit-evasion process with respect to different parameters. Right: The corresponding escape and capture results. (a) Results with respect to the ranges of $a_e$ and $\Delta d_0$, given $a_p=0.6$. (b) Results with respect to the ranges of $a_e$ and $\Delta \theta_0$, given $a_p=0.6$. (c) Results with respect to the ranges of initial conditions $a_e$ and $a_p$. }
	\label{param456}
\end{figure}

\textbf{Effects of the centripetal acceleration $r$}

As explained in Table~\ref{notations}, $r$ represents the centripetal acceleration, and a larger $r$ corresponds to faster turning. This makes $r$ one of the most important factors influencing the result. Fig.~\ref{param123}(a) shows the results as $r_e$ and $r_p$ vary. As can be seen from the result, the evader is more likely to be caught when $r_p$ is small (area \ding{172}) or $r_e$ is small (area \ding{173}). In area \ding{173}, since the pursuer is chasing behind the evader, the evader is more likely to be caught with a slower turning (see Fig.~\ref{capture}(a)). However, the possibility of capture decreases when $r_p$ is sufficiently large, as shown in both figures of Fig.~\ref{param123}(a). This is because excessive turning by the pursuer can cause it to overshoot and turn back before intercepting the evader. Even though $r_e$ is higher in area \ding{172}, allowing the evader to turn faster, it may meet the pursuer while turning back (see Fig.~\ref{capture}(b)). Nevertheless, it still makes sense to choose relatively large $r_e$. 
The same conclusion can also be obtained from (b) which shows the impacts of $r_e$ and $c_p$. 
The L-shaped capture area indicates that $c_p$ decreases as $r_e$ increases, implying that to capture an evader with faster turning, the pursuer needs to greatly decrease the velocity to increase the centripetal acceleration.

\textbf{Effects of the initial conditions $\Delta d_0$ and $\Delta \theta_0$}

Fig.~\ref{param123}(c) illustrates the capture and escape outcomes under different initial conditions $\Delta d_0$ and $\Delta \theta_0$ ($\Delta \theta_0=\theta_p(t_0)-\theta_e(t_0)$). On the one hand, capture is more likely to occur when the initial relative distance is short, with the probability decreasing as $\Delta d_0$ increases. This fact suggests the pursuer to get close to the evader to within a reasonable range (around 2-2.5 meters in this example). Conversely, the evader should flee immediately upon noticing the pursuer. That is why predators usually approach close enough to launch a successful attack (Section 1 of \cite{owen2019ramifying}).
On the other hand, capture tends to occur within the range $\Delta \theta_0\in [\pi/2,\pi]$ and $\Delta \theta_0\in [-\pi,-\pi/2]$. It means that it is easier to capture the evader when it initially faces the pursuer, as the evader needs more time to turn and escape in such cases.

\textbf{Effects of the tangential acceleration $a$}

The results with respect to the tangential acceleration $a$ are shown in Fig.~\ref{param456}(c). In area \ding{172}, the evader has larger deceleration, enabling it to perform quicker turning maneuvers (see Fig. ~\ref{escape}(c)). In area \ding{173}, the pursuer turns slightly quicker than the evader 
(see Fig.~\ref{escape}(b)). As the deceleration of the pursuer becomes larger, it turns much quicker than the evader at the beginning of taking turning maneuvers. This is why the evader escapes in area \ding{174} (see Figs.~\ref{escape}(a)(d)). The result shares the same conclusion with \cite{wilson2018biomechanics} that escaping at lower speeds allows prey to use their maximum manoeuvering performance and favour prey survival.

In contrary, the evader is caught in areas \ding{175}, \ding{176}, and \ding{177}. In area \ding{175}, both the pursuer and evader exhibit relatively smaller deceleration. The evader is captured when they both decelerate (see Fig.~\ref{capture}(a)). In area \ding{176} where the pursuer has a large acceleration while the evader has a very small one. The evader is caught when the pursuer finishes deceleration and accelerates (see Fig.~\ref{capture}(c)). Even when the evader turns faster at the beginning due to higher deceleration, it can still be caught as it accelerates (see Fig.~\ref{capture}(b)). This is why the evader is captured in area \ding{177}. 

In Figs.~\ref{param456} (a) and (b), $a_p=0.6$ while $a_e$ varies. The settings correspond to that of the last column of the right figure in (c), and they share the same escape area.
In summary, the result in Fig.~\ref{param456} suggests the evader to use smaller acceleration when $a_p$ is small, and use larger acceleration when $a_p$ is large.

\textbf{Effects of the alert distance $\varepsilon_1$}

\begin{figure}
	\centering
	\includegraphics[width=0.48\textwidth]{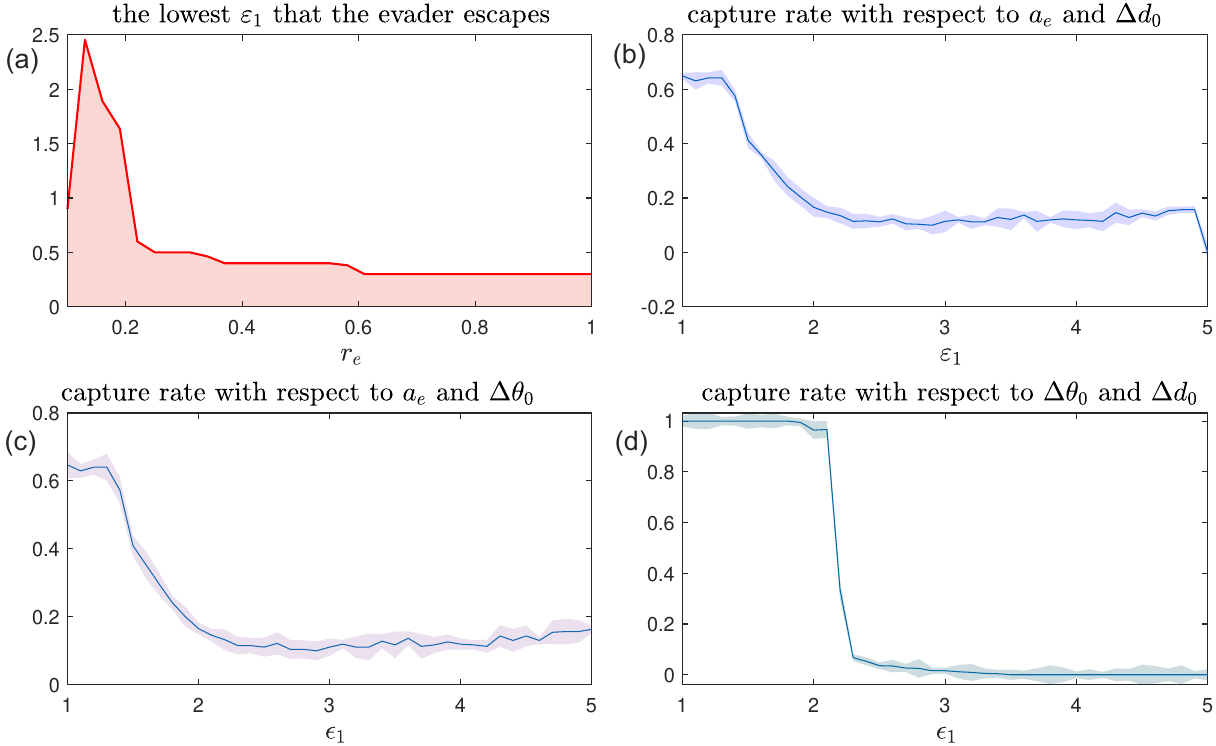}
	\caption{Escape results with respect to $r_e$ and  $\varepsilon_1$. Figure (a) shows the variation of the lowest alert distance $\varepsilon_1$ that the evader escapes with $r_e$. Each value is the mean of $\varepsilon_1$ under which the evader escapes given a range of the initial distance $\Delta d_0$. Figures (b)--(d) show the changes of capture rate with $\varepsilon_1$. Given the value of $\varepsilon_1$, the rate is the ratio of capture to all cases with respect to $a_e$ and $\Delta d_0$, $a_e$ and $\Delta \theta_0$, and $\Delta \theta_0$ and $\Delta d_0$, respectively. }
	\label{param7}
\end{figure}

As concluded from Fig.~\ref{param123}, a large $r_e$ is profit for the escape of evaders, and such observation is further confirmed by Fig.~\ref{param7}(a) that illustrates how the lowest alert distance $\varepsilon_1$, under which the evader successfully escapes, varies with $r_e$. Here, to provide a more accurate reflection of the trend, we use the average of the lowest $\varepsilon_1$ across a series of initial distance $\Delta d_0$.
Specifically, for each $r_e$, the lowest $\varepsilon_1$ is calculated as
\begin{equation}
	\frac{1}{m}\sum_{i=1}^m {\min}\{\mathop{arg}_{\varepsilon_1}{\min}\{\Delta d(\varepsilon_1,\Delta d_{i,0},t_f)\le \varepsilon_2\}\}, \label{low}
\end{equation}
where $m$ is the number of initial distance $\Delta d_0$. The relative distance between the pursuer and evader $\Delta d(\cdot,t)$ over time $t$ is a function of all parameters in Table~\ref{notations}. Here we simplify the notation in (\ref{low}) by assuming other parameters unchanged to emphasize the influence of $\varepsilon_1$ and $\Delta d_{i,0}$.
It can be seen that the lowest $\varepsilon_1$ decreases as $r_e$ increases. 
However, it is worth noting that $\varepsilon_1$ does not change as $r_e$ is large enough (greater than 0.6 in this example). This is because the angular velocity $w$ depends not only on the centripetal acceleration $r$ but is also limited by the maximum value $W^{\max}$ (see Eqs.~(\ref{evd_long}), (\ref{pur_long}) and (\ref{w-short})). 

The alert distance $\varepsilon_1$ is another one of the most important factors in the Alert-Turn algorithm. As shown in Figs.~\ref{param7} (b)--(d), the capture rate decreases significantly at first as $\varepsilon_1$ increases, and then the curves flatten as $\varepsilon_1$ continues to increase (Figs.~\ref{param7}(b)--(d)). 
The results suggest the evader to escape as early as possible once noticing the pursuer.
The capture rates are calculated as follows.

Let $\mathcal{A}$ denote the set of finite values of $a_e$, $\mathcal{D}$ the set of finite initial conditions $\Delta d_0$, and $\mathcal{T}$ the set of finite initial relative orientations $\Delta \theta_0$. The capture rate shown in Fig.~\ref{param7}(b) is obtained for each $\varepsilon_1$ by
\begin{align}
	&{\rm {rate}_1} = \frac{|\mathcal{S}_1|}{|\mathcal{A}||\mathcal{D}|},   \nonumber\\
	&\mathcal{S}_1 \!=\!\{(a_e,\Delta d_0) | \min \{\Delta d(a_e,\Delta d_0,t_f)\} \!\le\! \varepsilon_2,a_e \!\in\! \mathcal{A}, \Delta d_0 \!\in\! \mathcal{D}\}, \label{rate1}
\end{align}
where $|\mathcal{S}|$ denotes the cardinality of the set $\mathcal{S}$. The set $\mathcal{S}_1$ denotes the set of parameters that lead to a successful capture.

Similarly, the capture rates in Figs.~\ref{param7}(c)--(d) are, respectively, defined as, for each $\varepsilon_1$,
\begin{align}
	&{\rm rate_2} = \frac{\mathcal{S}_2}{|\mathcal{A}||\mathcal{T}|},   \nonumber\\
	&\mathcal{S}_2 \!=\! \{(a_e,\Delta \theta_0) | \min \{\Delta d(a_e,\Delta \theta_0,t_f)\}\!\le\! \varepsilon_2,a_e \!\in\! \mathcal{A}, \Delta \theta_0 \!\in\! \mathcal{T}\},  \nonumber  \\
	&{\rm rate_3} = \frac{\mathcal{S}_3}{|\mathcal{D}||\mathcal{T}|},\nonumber\\
	&\mathcal{S}_3 \!=\! \{(\Delta d_0,\Delta \theta_0) | \min \{\Delta d(\Delta d_0,\Delta \theta_0,t_f)\} \!\le\! \varepsilon_2,  \nonumber\\
 & \qquad \qquad \qquad \qquad \qquad \qquad \Delta d_0 \!\in\! \mathcal{D}, \Delta \theta_0 \!\in\! \mathcal{T}\}. \nonumber
\end{align}

\section{Pursuit-evasion Involving Multiple Pursuers and Multiple Evaders} \label{sec4}
The above algorithm is designed mainly for solitary animals. In what follows, the algorithm is extended to address the pursuit-evasion problem involving multiple robots, representing social animals that typically act in groups. In nature, preys usually perform different escape behavior patterns under the predator's attack, according to the habits of their species, and preys of the same species even do not respond to the pursuer's attack in the same way \cite{lee2006dynamics}. Research has identified two basic escape patterns: cooperative escape and selfish escape, and also their combinations \cite{lee2006dynamics,li2024intelligent}. Usually, anti-predator benefits increase with group size, and pursuers also cooperate to improve their chances of capturing prey\cite{jackson2006evolution}.

The whole algorithm is designed based on the Alert-Turn strategy. The control strategies, detailed in Section~\ref{sec4.1}, allow pursuers to set the number of targets based on the number of evaders. Pursuers cooperate to select their own targeted evader, which may involve choosing the same or different evaders. Once an evader is captured, it stops escaping, and the pursuer that captures this evader either stop or change its target \cite{jackson2006evolution}. The pursuers that did not capture the target but were chasing it will change their targets and continue pursuing.
Moreover, the pursuers are allowed to detect targets every $\Delta \bar t$ time interval. A new target is set if the predicted capture time is less than $pt$ times that of the previous target, that is, the evader $new\_tg$ is a new target if $t_{new\_tg}<pt\cdot t_{old\_tg}$ in which $t_{new\_tg}$ and $t_{old\_tg}$ are the predicted capture times, and $old\_tg$ is the previous target. In Section~\ref{sec4.3}, we present and analyze the influence of the escape patterns of the evaders, the number of pursuers, $\Delta \bar t$, and $pt$. The whole algorithm is shown in Algorithm~\ref{alg_ATA_m}. Note that $N_p$, $N_e$, $N_t$ and $N_{cp}$ denote, respectively, the number of the pursuers, the number of evaders, the number of targets to be captured, and the number of evaders that have been captured. $evd\_cap\_list$ represents the evaders that have been captured. $\mathcal{G}_m$ is the set of evaders belonging to the main group.

\begin{algorithm}
    \caption{Algorithm for the pursuit-evasion with multiple pursuers and multiple evaders }
\begin{algorithmic}[1] \label{alg_ATA_m}
    \STATE {\textbf{Input} $N_p$, $N_e$, $N_t$, $N_{cp}=0$, $\Delta \bar t$, $pt$, $evd\_cap\_list=[]$, $\mathcal{G}_m$}
    \WHILE{$N_{cp}<N_t$ \COMMENT{\%not all targets are captured}}
        \STATE{the pursuers selects targets every $\Delta \bar t$ seconds}
        \IF{$N_{cp}=N_t$}
            \STATE{all pursuers and evaders stop moving}
        \ELSE
            \FOR{$i=1:N_e$}
                \IF{ismember($i$,$evd\_cap\_list$)} 
                    \STATE{$v_{e,i}=0$, $w_{e,i}=0$}
                \ELSE
                    \IF{$\Delta d_{e,i,p}\ge \varepsilon_1$ \COMMENT{\%in long distance}}
                        \IF{$i\in \mathcal{G}_m$ \COMMENT{\%in the main group}}
                            \STATE{the evader $i$ executes the strategy (\ref{tot1}), (\ref{agg}), (\ref{md1})}
                        \ELSE
                            \STATE{the evader $i$ executes the strategy (\ref{tot2}), (\ref{md5}), (\ref{md2})}
                        \ENDIF
                    \STATE{The pursuers whose target is the evader $i$, execute strategy (\ref{vp_long})}
                    \ELSIF{$ \varepsilon_2<\Delta d_{e,i,p}\le  \varepsilon_1$ \COMMENT{\%in short distance}}
                        \STATE{the evader $i$ executes the strategy (\ref{md3})}
                        \STATE{the pursuers whose target is the evader $i$ execute strategy (\ref{md4})}
                    \ELSE
                        \STATE{$v_{e,i}=0$, $w_{e,i}=0$}
                        \STATE{update $N_{cp}$, $evd\_cap\_list$}                        
                    \ENDIF                    
                \ENDIF
                \STATE{update $\mathcal{G}_m$ \COMMENT{\%update the main group members}}
            \ENDFOR
        \ENDIF
    \ENDWHILE
\end{algorithmic}
\end{algorithm}

\subsection{Control Strategies for Players}  \label{sec4.1}
In our design, the evaders are divided into two groups: the main group and the isolated group. An evader is considered ‘isolated’ if it falls behind the other evaders after executing a turning maneuver. It re-joins the main group if its distance to the center of the remaining evaders exceeds a predefined threshold. 
Here, the center of a group is defined as $[x_c,y_c]^{\rm T}=\frac{1}{N}[x_1+x_2+\cdots+x_N,y_1+y_2+\cdots+y_N]^{\rm T}$ where $(x_i,y_i)$ denote the position of the $i$th player.
Those non-isolated evaders form the main group. In our method, the evaders in the isolated group and the main group execute different strategies. In particular, the velocity command for the $i$th evader during the \textbf{long-distance phase} is
\begin{align}
	v_{e,i,m} &= (1-\alpha) v_{e,i,m}^{agg} + \alpha v_{e,i,m}^{esc}, \nonumber\\
	  w_{e,i,m} &= (1-\alpha) w_{e,i,m}^{agg} + \alpha w_{e,i,m}^{esc},  \label{tot1} \\
	v_{e,i,s} &= (1-\beta(\Delta d_{pi})) v_{e,i,s}^{joi} + \beta(\Delta d_{pi}) v_{e,i,s}^{esc}, \nonumber\\
	w_{e,i,s} &= (1-\beta(\Delta d_{pi})) w_{e,i,s}^{joi} + \beta(\Delta d_{pi})w_{e,i,s}^{esc},  \label{tot2} 
\end{align}
where the subscripts $m$ and $s$ denote the main and isolated group, respectively. 
$v_{e,i,m}^{agg}$ and $w_{e,i,m}^{agg}$ represent the aggregation/formation velocity commands which enable the evader group to move in formation, $v_{e,i,m}^{esc}$ and $w_{e,i,m}^{esc}$ are the escape velocity commands. 
$0\le \alpha \le 1$ is a constant denoting the weight of the escaping term, and is referred to as the ‘selfish parameter'. The design of (\ref{tot1}) is to replicate the evaders' selfish escape pattern and collective escape pattern. It will further be discussed in Sections~\ref{sec4.2} and \ref{sec4.3}. 
For the isolated evaders, the strategy (\ref{tot2}) drives them to move towards the main group while simultaneously escaping from pursuers.
$0\le \beta(\Delta d_{pi})\le 1$ is a decreasing function of $\Delta d_{pi}$, the relative distance between the evader and the pursuer. The reason for this design is that as $\Delta d_{pi}$ increases, the pursuer has decreasing influence on the isolated evader, so the evader puts more effort into keeping up with the main group. 

The aggregation/formation velocity commands are designed as
 \begin{equation}  \label{agg}
 	\begin{aligned}
 		v_{e,i,m}^{agg} & \!\!=\! -(1\!-\!\sigma_{ii})\big[(x_{i,m}\!-\!x_{c})cos(\theta_{i,m})\!+\!(y_{i,m}\!-\!y_{c})sin(\theta_{i,m})\big]  \\
 		&~~+ V_{e,i,m}^{\max},  \\
 		w_{e,i,m}^{agg} &\!\!=\!\! -[\theta_{e,i,m} \!-\! (1-\sigma_{ii})\Delta \theta_{ic}-\sigma_{ii}\Delta \theta_{pc}], \\
 		\sigma_{ii} &= \frac{2}{1+e^{-((d_{ic}^{des})^2-\Delta d_{ic}^2)}}. 
 	\end{aligned}
 \end{equation}
Here $x_{c}$ and $y_c$ are the $x$ and $y$ position, respectively, of the center of the group composed of the evader $i$ and its evader neighbors. $d_{i,c}^{des}$ denotes the desired distance between the $i$th evader and the center of the group, and it is updated at a certain frequency. $\Delta \theta_{ic}$ and $\Delta \theta_{pc}$ represent the relative angle between the $i$th evader and the center, and between the pursuer and the center, respectively, given by ${\rm atan}\frac{y_c-y_i}{x_c-x_i}$ and ${\rm atan}\frac{y_c-y_p}{x_c-x_p}$, respectively.

Denote $q_{di}=[x_{i,m}-x_c \quad y_{i,m}-y_c]^{\rm T}$, it is obvious that $\|q_{di}\|=\Delta d_{ic}$. 
By \cite{gazi2004class}, it satisfies that
\begin{equation}
    \dot q_{di}\!=\!-q_{di}(1\!-\!\sigma_{ii})=\!-q_{di}(1\!-\frac{2}{1+e^{\|q_{di}\|^2-(d_{ic}^{des})^2}}).  \label{qdi}
\end{equation}
From \eqref{qdi} we can draw the following conclusion.
\begin{theorem} \label{them2}
Let $\Omega_e=\{q_{di}:\|q_{di}\|=d_{ic}^{des}\}$ denote the invariant set of equilibrium points. Then, 
$ q_{di}\to \Omega_e$ as $t\to \infty$ if $\|q_{di}(t_0)\|>0$.
\end{theorem}
The proof is given in the appendix. The theorem means that the swarm converges to a configured aggregation clearance distance $d_{ic}^{des}$. 
The aggregation velocity commands (\ref{agg}) aim to drive evaders in the main group to aggregate at configurable clearance distance. $1-\sigma_{ii}$ and $\sigma_{ii}$ are used in $w_{i,m}^{agg}$ to ensure that the evaders align as they converge to the configured aggregation clearance distance and stay away from the pursuer. 

The escape velocity commands are designed as
\begin{equation} \label{md1}
	 \begin{aligned}
	v_{e,i,m}^{esp} &= v_{e,i,m}(\bar t_{e,i,m}) + a_{e,i,m} (t-\bar t_{e,i,m}),\\
        w_{e,i,m}^{esp} &= -(\theta_{e,i,m}-\Delta \theta_{pi}).
	 \end{aligned}
\end{equation}

Note that the term $\Delta \theta_{pc}$ in (\ref{agg}) will help the evaders in the main group flee from the pursuer, even in the absence of an escape command, i.e., $\alpha=0$. However, this does not mean that the escape command is meaningless. The weight $\alpha$ determines the degree of evaders' dispersion during escaping. A larger $\alpha$ results in a greater degree of dispersion, as Fig.~\ref{dispersion} shows.

The evaders in the \textit{isolated group} split from the main group, and they definitely try to stay away from the pursuer, and at the same time, to keep up with the main group to seek protection. In such a case, the commands for the evaders in the isolated group also consist of two parts: rejoining and escaping.

Specifically, the rejoining objective is achieved by driving them to move closer to the center of the main group, by the following velocity command
\begin{equation} \label{md5}
\begin{aligned}
    v_{e,i,s}^{joi} &= v_{e,i,s}(\bar t_{e,i,s})+a_{e,i,s} (t-\bar t_{e,i,s}), \\
    w_{e,i,s}^{joi} &= -(\theta_{e,i,s}-\Delta \theta_{ic}).
\end{aligned}
\end{equation}
The escape velocity command is
\begin{equation}\label{md2}
\begin{aligned}
    v_{e,i,s}^{esp} &= v_{e,i,s}(\bar t_{e,i,s}) + a_{e,i,s} (t-\bar t_{e,i,s}), \\
    w_{e,i,s}^{esp} &= -(\theta_{e,i,s}-\Delta \theta_{pi}).
\end{aligned}
\end{equation}
 
Here, pursuers may select the same target or different targets, but they follow the same target selection principle (the closed distance or the shortest time). We denote the $j$-th pursuer targeting evader $i$ as $p_j(e_i)$ (or simply $p_j$), and the closest pursuer to evader $i$ as $p_l(e_i)$ (or simply $p_l$).
During the long-distance phase, pursuers take inter-player collision avoidance strategies while chasing their targets. Therefore, the velocity commands for the pursuers consist of two parts. Specifically, the commands are designed as

\begin{subequations}\label{vp_long}
    \begin{align} 
		&v_{p_j} = \left\{ 
		\begin{aligned}
			&{\rm sat}(v_{p_j}(\bar t_{p_j})+a_{p_j} (t-\bar t_{p_j}), V_{p_j}^{\max}),\\ & \qquad\qquad\qquad\quad {\rm if}~\|\theta_{p_j}-\Delta \theta_{p_j,e_i}\|\leq \bar \theta \\
			&{\rm sat2}(v_{p_j}(\bar t_{p_j})-a_{p_j}(t-\bar t_{p_j}), c_{p_j}V_{p_j}^{\max}),\\ & \qquad\qquad\qquad\qquad\qquad  \quad  {\rm otherwise} , 
		\end{aligned}
		\right.  \\
		&w_{p_j} = -{\rm sgn}\big(\theta_{p_j}-\Delta \theta_{p_j,e_i}-\sum_{k=1}^{N_p}rep_{p_j,k},  \label{YYb}\\
		& \qquad \qquad\qquad \qquad \min (r_{p_j} v_{p_j}^{-1},W_{p_j}^{\max})\big), \\
		&rep_{p_j,k} = \left\{ 
		\begin{aligned}
			&\frac{m}{\Delta d_{p_j,k}}\Delta \theta_{p_j,k}, &{\rm if}\Delta d_{p_j,k}<d_{safe} \\
			& 0, & {\rm otherwise}
		\end{aligned}
		\right.
	\end{align}
\end{subequations}
where $rep_{p_j,k}$ denotes the repelling force among pursuers, and the force becomes larger as the distance decreases. Collision avoidance is achieved by changing the orientation of pursuers, as (\ref{YYb}) shows.

\begin{figure*}	
	\centering
	\includegraphics[width=0.245\textwidth]{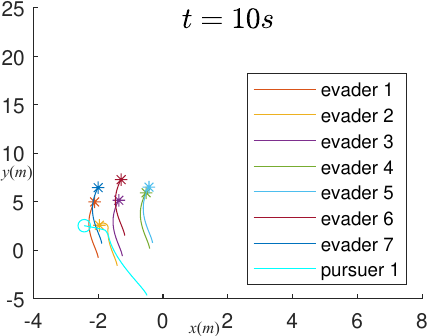}
	\includegraphics[width=0.245\textwidth]{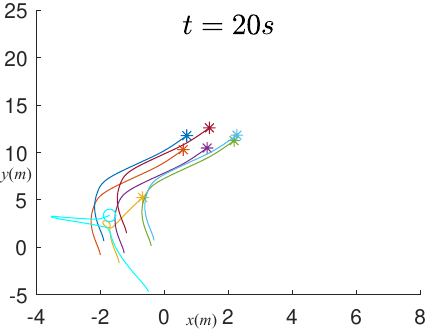}
	\includegraphics[width=0.245\textwidth]{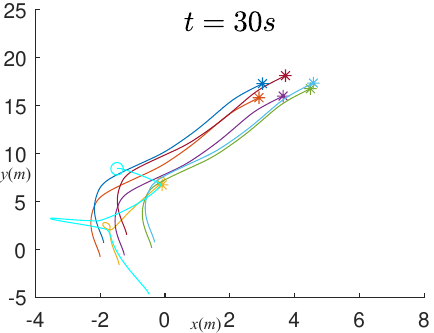}
	\includegraphics[width=0.245\textwidth]{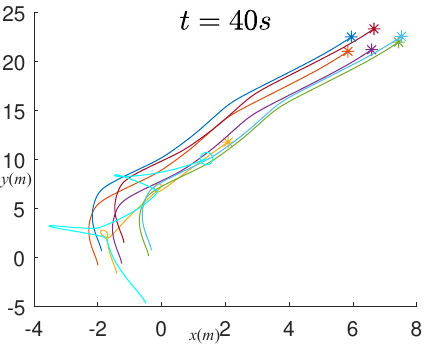}
	\caption{Illustration of one motion pattern with one pursuer and seven evaders. $\alpha=0$, $\Delta \bar t=\infty$, $\epsilon_1=1$, $\epsilon_2=0.1$, $V_p^{\max}=1$, $W_p^{\max}=1$, $r_p=0.1$, $c_p=0.05$, $a_p=0.6$, and for each evader, $V_e^{\max}=0.6$, $W_e^{\max}=2$, $r_e=0.2$, $c_e=0.05$, $a_e=0.3$.}
	\label{formation}
\end{figure*}

\begin{figure*}
	\centering
	\includegraphics[width=0.271\textwidth]{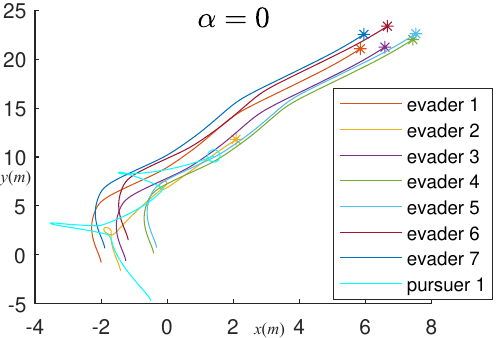}
	\includegraphics[width=0.231\textwidth]{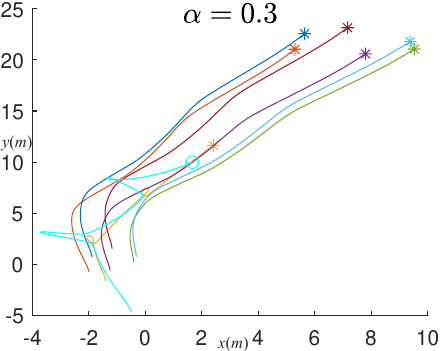}
	\includegraphics[width=0.231\textwidth]{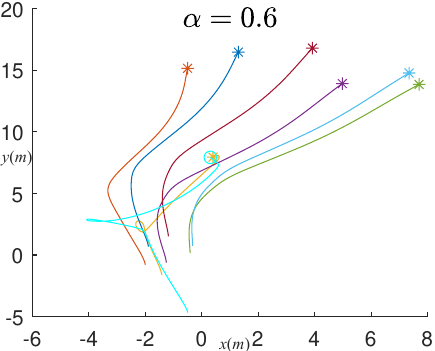}
	\includegraphics[width=0.231\textwidth]{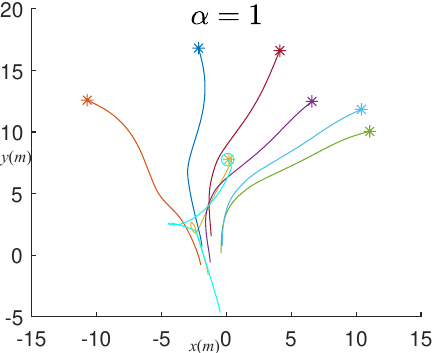}
	\caption{Illustration of motion patterns with respect to different value of the selfish parameter $\alpha$. $t_f=40$, $\Delta \bar t=\infty$, $\epsilon_1=1$, $\epsilon_2=0.1$, $V_p^{\max}=1$, $W_p^{\max}=1$, $r_p=0.1$, $c_p=0.05$, $a_p=0.6$, and for each evader, $V_e^{\max}=0.6$, $W_e^{\max}=2$, $r_e=0.2$, $c_e=0.05$, $a_e=0.3$.}
	\label{alpha}
\end{figure*}

The evader $i$ is considered to enter the \textbf{short-distance phase} when its distance to the closed pursuer $p_l(e_i)$ is less than $\varepsilon_1$. During this phase, both the evader and the pursuers $p_j(e_i)$ take turning maneuvers. Specifically, the strategy of the evader is
\begin{equation} \label{md3}
	\begin{aligned}
		v_{e,i} &\!=\! \left\{ 
		\begin{aligned}
			&{\rm sat}\big(\max\{0,v_{e,i}(\bar t_{e,i})\!-\! a_{e,i} (t-\bar t_{e,i})\}, V_{e,i}^{\max}\big), \\& \qquad \qquad \qquad \qquad \qquad \qquad ~~{\rm if}~ v_{e,i} \!>\! c_{e,i} V_{e,i}^{\max} \\
			& {\rm sat}(v_{e,i}(\bar t_{e,i})\!+\!a_{e,i} (t-\bar t_{e,i}), V_{e,i}^{\max}),~~{\rm otherwise},
	    \end{aligned}
    \right.  \\
    	w_{e,i} &\!=\! \left\{ 
    	\begin{aligned}
    		&{\rm sat}(-{\rm acot}(k_e \Delta d_{pe_i})v_{e,i}^{-1},\min \{r_{e,i}v_{e,i}^{-1},W_{e,i}^{\max}\}), \\& \qquad \qquad \qquad \qquad \qquad \qquad ~~{\rm if}~\theta_{p_l}-\theta_{e_i}>0 \\
    		&{\rm sat}({\rm acot}(k_e \Delta d_{pe_i})v_{e,i}^{-1},\min \{r_{e,i}v_{e,i}^{-1},W_{e,i}^{\max}\}), \\& \qquad \qquad \qquad \qquad \qquad \qquad \qquad  {\rm otherwise.}
    	\end{aligned}
    \right.  \\
	\end{aligned}
\end{equation}
The strategy of the pursuer follows
\begin{equation} \label{md4}
	\begin{aligned}
		v_{p_j} &\!=\! \left\{ 
		\begin{aligned}
			&{\rm sat}\big(\max\{0,v_{p_j}(\bar t_{p_j})\!-\!a_{p_j} (t-\bar t_{p_j}))\}, V_{p_j}^{\max}\big), \\& \qquad \qquad \qquad\qquad \qquad  {\rm if}~ v_{p_j} \!>\! c_{p_j}  V_{p_j}^{\max} \\
			& {\rm sat}(v_{p_j}(\bar t_{p_j})\!+\!a_{p_j}(t-\bar t_{p_j})), V_{p_j}^{\max}), ~~ {\rm otherwise},
		\end{aligned}
		\right.  \\
        w_{p_j} &\!=\!{\rm sign}(w_{e,i}){\rm sat}({\rm acot}(k_p \Delta d_{pe_i})v_{p_j}^{-1},\min \{r_{p_j}v_{p_j}^{-1},W_{p_j}^{\max}\}).
	\end{aligned}
\end{equation}
For the evader $i$ and the pursuers $p_j$, the short-term strategies (\ref{md3}) (\ref{md4}) are the same as (\ref{v-short}) (\ref{w-short}). 

When a target is captured, it will stop without a doubt. 

When there are multiple pursers and one evader, the evader takes only the escape strategy (\ref{md1}) or (\ref{md2}), without the aggregation term, and the pursuers chase the unique evader.  All players stop if the evader is captured. When there are multiple evaders and one pursuer, the repelling force in (\ref{vp_long}) does not exist anymore. Thus, the command simplifies to
\begin{equation*}
    	\begin{aligned}
		v_{p} &= \left\{ 
		\begin{aligned}
			&{\rm sat}(v_{p}+a_{p}(t-\bar t_p), V_{p}^{\max}), ~ {\rm if}~\|\theta_{p}-\Delta \theta_{p,e_i}\|\leq \bar \theta \\
			&{\rm sat2}(v_{p}-a_{p}(t-\bar t_p), c_{p}V_{p}^{\max}), ~ {\rm otherwise}  ,
		\end{aligned}
		\right.\\
		w_{p} &= -{\rm sgn}\big(\theta_{p}-\Delta \theta_{p,e_i}, \min (r_{p} v_{p}^{-1},W_{p}^{\max})\big). \\
		\end{aligned}
\end{equation*} 
The pursuer can also change targeted evaders, and the evader stops moving once captured.

\begin{remark}
    The pursuit-evasion strategy for multiple pursuers and multiple evaders combines the basic Alert-Turn strategy designed in Section \ref{sec3.2} with cooperative terms that allow robots to collaborate within their respective teams. It can be seen that the long-distance strategy (\ref{md1}), (\ref{md2}) and (\ref{vp_long}) extend the strategy (\ref{evd_long}) and (\ref{pur_long}) of Section \ref{sec3.2}, and the short-term strategies (\ref{md3}) (\ref{md4}) are equivalent to the strategies (\ref{v-short}) (\ref{w-short}). 
\end{remark}

\subsection{Results with Respect to a Single Run} \label{sec4.2}
In this section, we present some motion patterns involving multiple robots to visualize the strategies of evaders and pursuers. 

Fig.~\ref{formation} shows a motion pattern involving one pursuer and seven evaders. In this example, the selfish parameter $\alpha=0$ indicates that the evaders escape cooperatively. We can see from the figures that the evaders of the main group (evaders 1 and 3-7) move in formation under the aggregation velocity commands (\ref{agg}). The expected formation configuration $d_{i,c}^{des}$ is defined based on their initial positions. The result is similar to some observations of escaping behaviors of prey that keep close with their groups \cite{yang2014aggregation}. The advantage is that once they get stuck, they have more power to fight against predators. In this example, the pursuer selects evader 2 as its target. After taking the turning maneuver, evader 2 falls behind the others, becoming the isolated evader, so it tries to catch up with the main group.

With the increasing value of the selfish parameter $\alpha$, evaders escape more selfishly (see the middle two figures of Fig.~\ref{alpha}). As $\alpha$ increases to 1, they flee completely selfishly, as illustrated in the last figure of Fig.~\ref{alpha}. This escape pattern is similar to that of antelope, zebra and so on. In Section~\ref{sec4.3}, a quantitative analysis of $\alpha$ is given.

\begin{figure}
	\centering
	\includegraphics[width=0.4\textwidth]{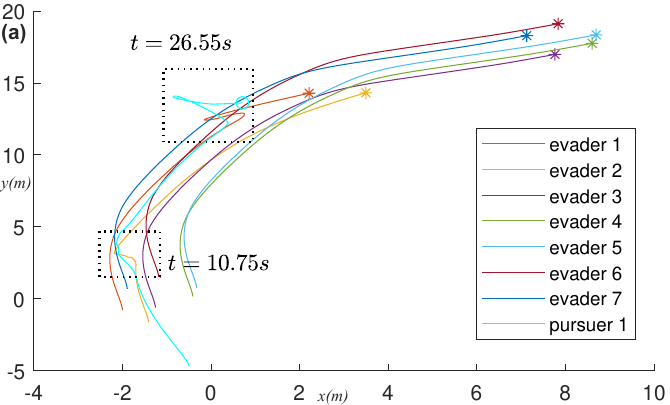}
	\includegraphics[width=0.4\textwidth]{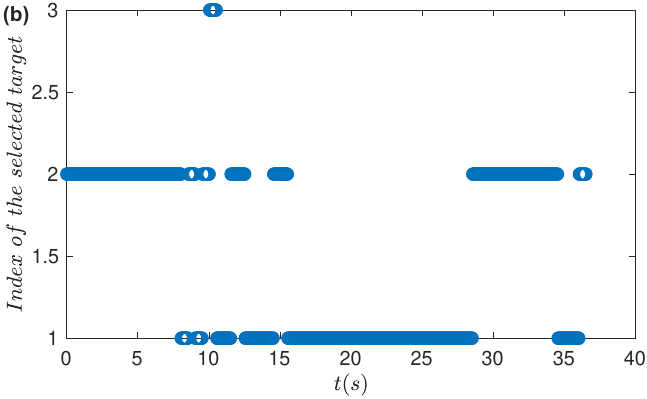}
	\caption{Illustration of one motion pattern with the pursuer using target changing mechanism. (a) Trajectories of agents. (b) Index of selected targets at different times. $\alpha=0$, $pt=0.2$, $\Delta \bar t=0.5$, $\epsilon_1=1$, $\epsilon_2=0.1$, $V_p^{\max}=1$, $W_p^{\max}=1$, $r_p=0.1$, $c_p=0.05$, $a_p=0.6$, and for each evader, $V_e^{\max}=0.6$, $W_e^{\max}=2$, $r_e=0.15$, $c_e=0.05$, $a_e=0.3$.}
	\label{targ}
\end{figure}

\begin{figure}
	\centering
	\includegraphics[width=0.38\textwidth]{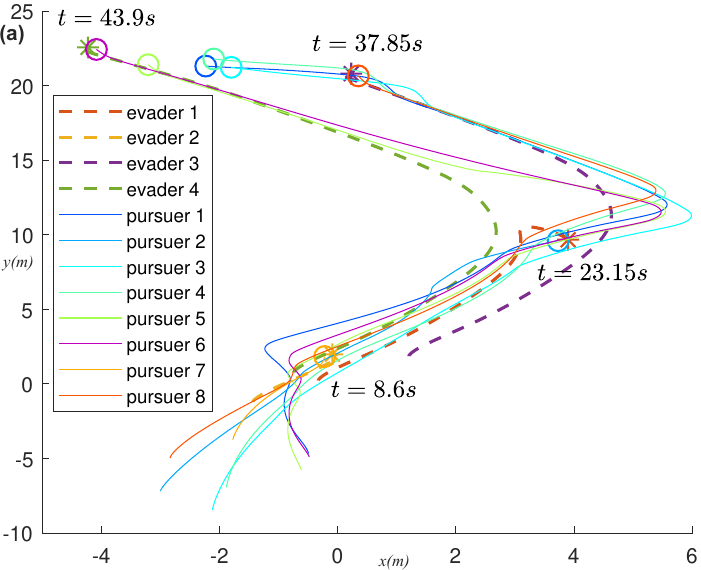}
	\includegraphics[width=0.109\textwidth]{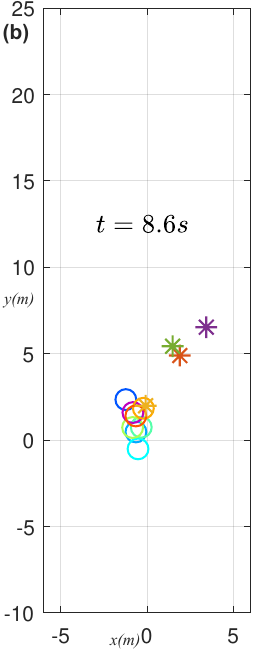}
	\includegraphics[width=0.132\textwidth]{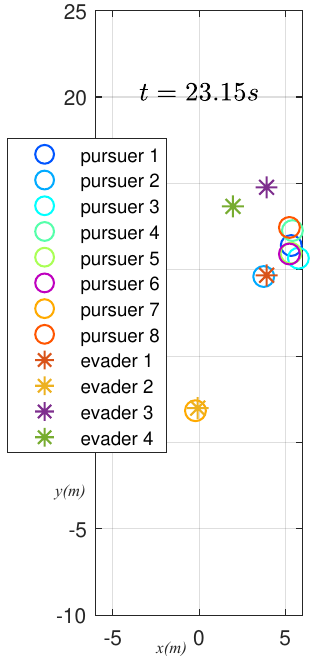}
	\includegraphics[width=0.108\textwidth]{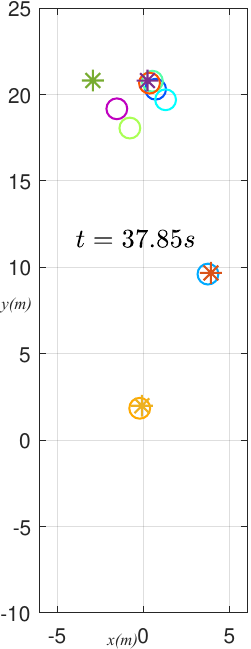}
	\includegraphics[width=0.11\textwidth]{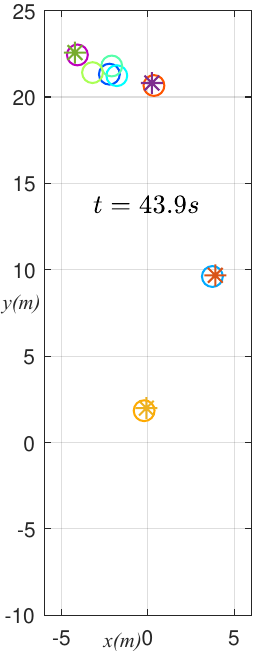}
	\caption{Illustration of one motion pattern involving four evaders and eight pursuers. (a) Trajectories of agents. (b) Positions of agents at the time each evader is captured. $\alpha=0$, $pt=0.2$, $\epsilon_1=1$, $\epsilon_2=0.2$. For each pursuer, $V_p^{\max}=1$, $W_p^{\max}=1$, $r_p=0.1$, $c_p=0.3$, $a_p=0.6$, and for each evader, $V_e^{\max}=0.6$, $W_e^{\max}=2$, $r_e=0.2$, $c_e=0.1$, $a_e=0.35$.}
	\label{MM}
\end{figure}

Fig.~\ref{targ} illustrates how systems evolve when the pursuer dynamically changes its target. As shown in (b), the pursuer selects evader 2 as its target at the beginning. At about $t=10.75 s$, evader 2 takes a turning maneuver and survives. Subsequently, evader 1 becomes the main target. At about $t=26.55s$, evader 1 also turns and survives. After that, both evaders 1 and 2 become isolated and try to rejoin the main group. In the following subsection, the influence of target detection and changing frequencies is given and analyzed. 

Fig.~\ref{MM} presents a motion pattern involving four evaders and eight pursuers, where the pursuers aim to capture all evaders, and each pursuer selects its target separately. Once an evader is caught, the pursuer who captures it will stop and other pursuers continue to chase the left evaders. Evaders 2 is firstly captured at $t=8.6 s$. When evader 1 is captured at about $t=23.15 s$, pursuers 5 and 6 try to catch evader 4, and pursuers 1, 3 and 4 target evader 3. Finally, evaders 3 and 4 are caught at $t=37.85 s$ and $t=43.9 s$. Due to the collision avoidance strategy of pursuers, they will not collide with each other. In this example, $d_{safe}=0.8$ (defined in (\ref{vp_long})). 

\subsection{Results with Respect to Parameters’ Changes}  \label{sec4.3}
In Section~\ref{sec3.4}, influences of the parameters of the Alert-Turn algorithm are given and analyzed. In this section, we focus on discussing key parameters that are specific to the pursuit-evasion problem involving multiple pursuers and evaders.

\begin{figure}
	\centering
	\includegraphics[width=0.39\textwidth]{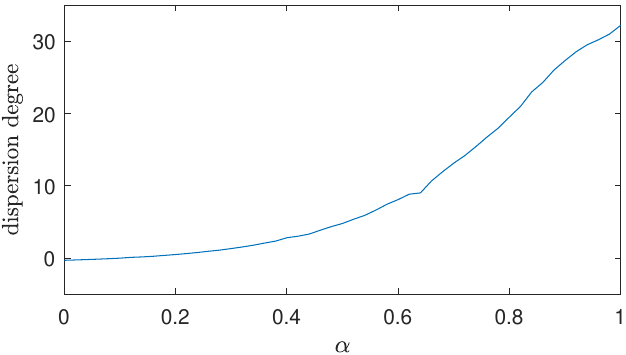}
	\caption{Illustration of dispersion degree with respect to the selfish parameter $\alpha$.}
	\label{dispersion}
\end{figure}

\begin{figure*}
	\centering
	\includegraphics[width=0.2468\textwidth]{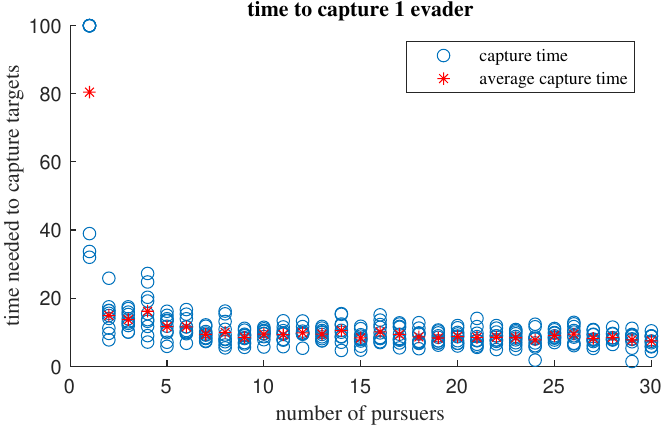}
	\includegraphics[width=0.2468\textwidth]{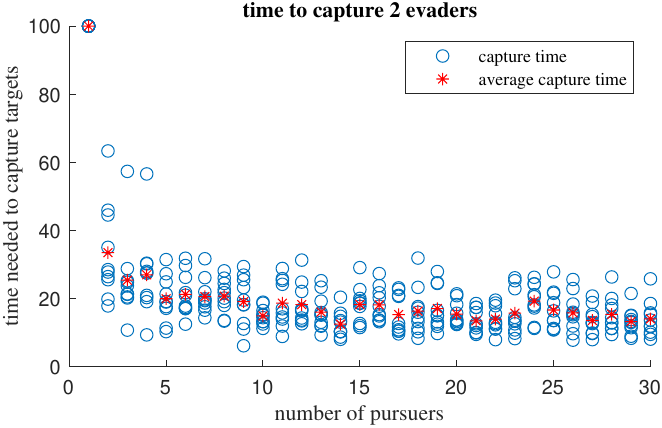}
	\includegraphics[width=0.2468\textwidth]{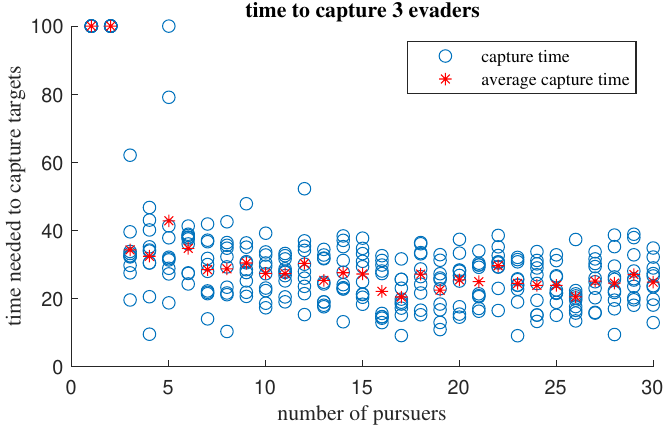}
	\includegraphics[width=0.2468\textwidth]{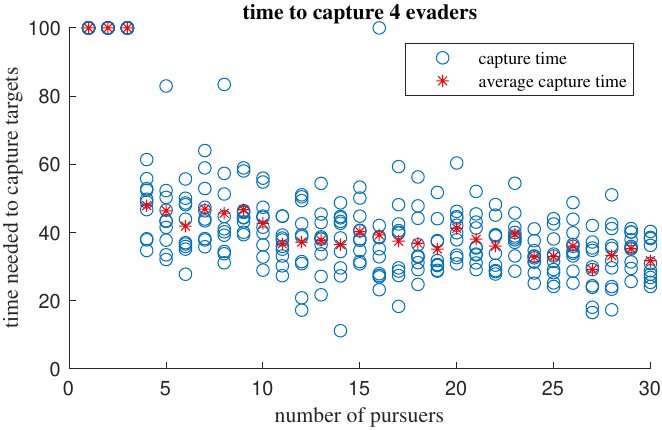}
	\caption{Time needed to capture 1, 2, 3, and 4 target(s), respectively, with respect to the number of pursuers. }
	\label{Np_targets}
\end{figure*}

Fig.~\ref{dispersion} illustrates the dispersion degree with respect to the selfish parameter $\alpha$. The dispersion degree is defined as 
\begin{align*}
	&{\rm dispersion}\\
	&=\frac{\sum_{i=1}^{N_e}(pos_i(t_f)\!-\!pos_c(t_f))-\sum_{i=1}^{N_e}(pos_i(t_0)\!-\!pos_c(t_0))}{\sum_{i=1}^{N_e}(pos_i(t_0)-pos_c(t_0))}
\end{align*}
where $t_0$ is the initial time, $t_f$ is the simulation time, $pos_i$ and $pos_c$ denote the position of the $i$-th evader and their group center, respectively. We note that when $\alpha=0$, the dispersion degree is nearly 0, indicating the evaders keep their initial formation configuration while escaping. The dispersion degree increases as $\alpha$ increases. The dispersion degree appears to exhibit an approximately exponential relationship with $\alpha$, as observed in Fig.~\ref{dispersion}.

The effects of the number of pursuers on the capture time are demonstrated in Fig.~\ref{Np_targets}. The results are obtained from a statistical perspective. In our settings, the number of pursuers ranges from 1 to 30, and 30 groups of initial positions are randomly assigned to all robots. The number of evaders is 4, and the pursuers can capture 1, 2, 3, or 4 target(s). Fig.~\ref{Np_targets} presents the capture time for different settings, along with the average capture time. It is obvious that it takes the pursuers longer to capture more evaders. We can also see from the four figures that the overall capture time decreases as the number of pursuers increases. However, as the number of pursuers increases, the capture time decreases more slowly. This is because pursuers must put more effort to inter-robot collision avoidance as their density increases. 

\begin{figure}
	\centering
	\includegraphics[width=0.32\textwidth]{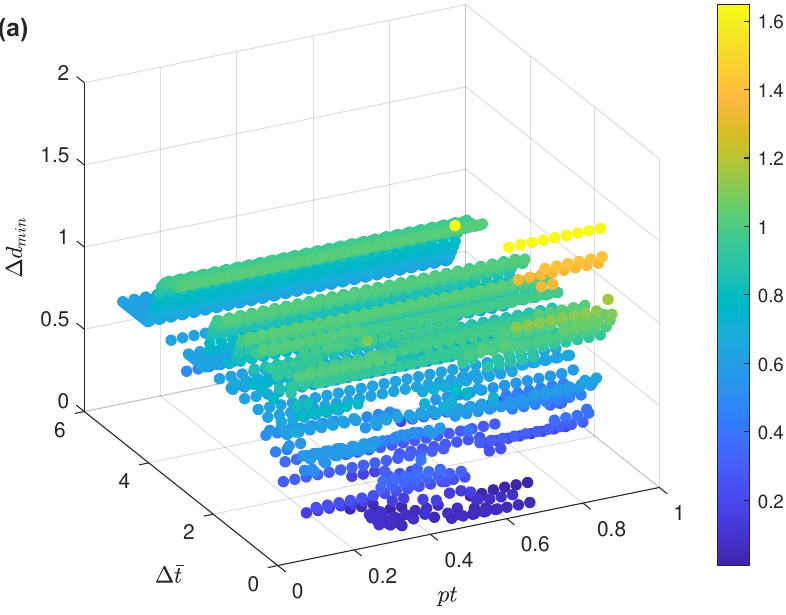}
        \includegraphics[width=0.32\textwidth]{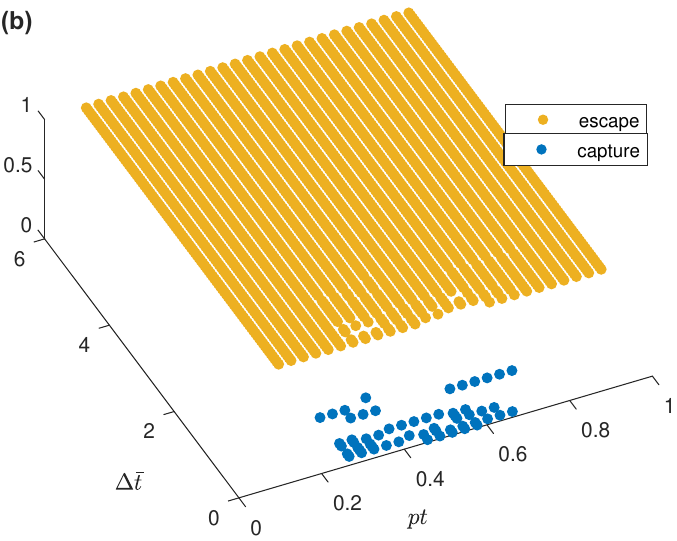}
	\caption{Simulation results with respect to the changes of $\Delta \bar t$ and $pt$. The numbers of the pursuer and the evaders are 1 and 10, respectively. $\varepsilon_1=1.4$, $\varepsilon_2=0.14$, $\alpha=0$, $V_p^{\max}=1.2$, $w_p^{\max}=1$, $r_p=0.1$, $a_p=0.6$, $c_p=0$, $t_f=100$. For each evader, $w_e^{\max}=2$, $r_e=0.2$, $a_e=0.3$, $c_e=0$. For evaders 4, 5, and 9, $V_{e,4}^{\max}=0.5$, $V_{e,5}^{\max}=0.9$, $V_{e,9}^{\max}=0.5$, while for the left evaders, $V_e^{\max}=0.6$.}
	\label{fre_pt}
\end{figure}

Fig.~\ref{fre_pt} shows the results under varying detection time interval $\Delta \bar t$ and target-changing thresholds $pt$. 
We can find that capture is predominantly achieved when $\Delta \bar t\in [0,1]$ and $pt\in [0.28,0.67]$ in our settings. This suggests that frequent detection of new targets is advantageous for pursuers. According to the definition of $pt$, a higher $pt$ value corresponds to more frequent target changes by the pursuer. The evaders escape as $pt>0.67$ or $pt<0.28$, indicating that both overly frequent and overly infrequent target changes by the pursuer negatively impact the likelihood of capture.

\section{Replication of Animals' Behaviours} \label{sec5}
In order to show the effectiveness of our algorithm, we aim to replicate some chasing and hunting behaviours of predators and preys in nature. To evaluate the similarity between real behaviors and our simulation results, we conduct comparisons focusing on relative distance and time.

Firstly, we use YOLOv8 
to detect five segments of wild animal chasing videos. The detection aims to locate the predator and prey in the videos and to analyze their relative positional changes during the chase. During the image pre-processing, we extract images from the videos at a rate of 10 frames per second, and use each frame as an input image. The output includes the bounding box coordinates and class information for each detected object, and we use the center of the bounding box to represent the position of each object. 

The comparison results can be found in the video https://youtu.be/VC9oyULhnlw,
which shows our strategy effectively replicates the chasing behaviors of \textit{various animal species} across \textit{different scenarios}. 
In the first two scenarios, two dogs chase one rabbit, while the last three scenarios show one cheetah chasing one deer. Given the positions of predators and prey, we calculate the relative distance between each pair of predator and prey.
It can be seen that our simulation results follow almost the same turning time and capture time, and similar trends of relative distance variation of each pair of the pursuer and the evader.


\section{Conclusion} \label{sec6}
In this paper, we have proposed a nature-inspired dynamic control framework for the pursuit-evasion of unicycle robots. The framework incorporates an Alert-Turn control strategy designed for scenarios with one pursuer and one evader, followed by the analysis of escape and capture results at a lower level of a single run and a higher level with respect to parameters’ changes. The strategy was extended by integrating aggregation control laws and a target-changing mechanism to model different pursuit and escape patterns. The effects of control laws and the mechanism were quantified statistically. It is notable that some findings are in line with observations in nature, which, together with the replication result, further validate the effectiveness of our strategies. In the near future, we plan to study the pursuit-evasion problem for robots navigating in obstacle environments and tackle the inverse optimal control problem, aiming to learn the objective functions of animals.

\section*{Appendix}
\textbf{Proof of Theorem~\ref{them1}}. Without loss of generality, we assume $t_0=0$, $x_p(0)=x_e(0)=0$, $y_p(0)=0$, and $y_e(0)=d_0$. 
Since $\Delta d(0)\le \varepsilon_1$, the evader and pursuer will take the short-phase strategy (\ref{v-short})--(\ref{w-short}). By (\ref{w-short}), since $\theta_e(0)=\pi/2$, the evader and pursuer turn left as $\theta_p(0)=\theta_0\in [\pi/2-\gamma,\pi/2]$, and turn right otherwise. The same result will be obtained as $\theta_0\in (\pi/2,3\pi/4)$. Thus, we only consider the case $\theta_0\in [\pi/2,\pi/2+\gamma]$ here.

Taking $k_e$, $k_p$, $r_e$ and $r_p$ of Theorem~\ref{them1} into (\ref{w-short}) gives $w_e=W_e^{\max}$ and $w_p=W_p^{\max}$, which leads to $\theta_p(t)=\theta_p+W_p^{\max}t$ and $\theta_e(t)=\pi/2+W_e^{\max}t$. 
For $t\in (0,\frac{\pi}{2W_p^{\max}})$, since $V_p-a_p\frac{\pi}{2W_p^{\max}} \ge c_pV_p^{\max}$ and $V_e-a_e\frac{\pi}{2W_p^{\max}} \ge c_eV_e^{\max}$, we have the linear velocity commands $v_p(t)=V_p-a_pt$ and $v_e(t)=V_e-a_et$. 
The above facts lead to the following dynamics of the pursuer and evader 
\begin{equation}
    \begin{aligned}
        \dot x_p&=(V_p-a_pt) cos(\pi/2-\gamma +W_p^{\max}t),\\
        \dot y_p&=(V_p-a_pt) sin(\pi/2-\gamma +W_p^{\max}t), \\
        \dot x_e&=(V_e-a_et) cos(\pi/2+W_e^{\max}t),\\
        \dot y_e&=(V_e-a_et) sin(\pi/2+W_e^{\max}t).  \label{mdd}
    \end{aligned}
\end{equation}

Since $\gamma$ is small enough by condition (v) of Theorem~\ref{them1}, it is easy to get $\dot x_p =- (V_p-a_pt)sin(W_p^{\max}t) + \delta_1(\gamma)$ with $\delta_1(\gamma)$ small enough and $\delta_1(\gamma)\to 0$ as $\gamma\to 0$. Similarly, we have $\dot y_p = (V_p-a_pt)sin(W_p^{\max}t) + \delta_2(\gamma)$ where $\delta_2(\gamma)$ has the same property as $\delta_1(\gamma)$.

Since $W_e^{\max}>W_p^{\max}$, $x_p>x_e$, and it is obvious that $y_e>y_p$ at the beginning. Therefore, the relative distance along the $x$-axis and $y$-axis are denoted by $\Delta x=x_p-x_e$ and $\Delta y=y_e-y_p$.  Denote $sin \alpha=\Delta x/\Delta d_{pe}$, the derivative of the relative distance of the pursuer and the evader is
\begin{equation*}
    \begin{aligned}
        &\dot {\Delta d_{pe}} =\frac{\Delta x}{\Delta d_{pe}}\dot {\Delta x} + \frac{\Delta y}{\Delta d_{pe}}\dot {\Delta y} \\
        &= sin \alpha[V_e sin(W_e^{\max}t)-V_psin(W_p^{\max}t) \\
        &\qquad \qquad -a_etsin(W_e^{\max}t)+a_ptsin(W_p^{\max}t) +\delta_1(\gamma)  ] \\
        &\quad + cos\alpha [ V_pcos(W_p^{\max}t)-V_ecos(W_e^{\max}t)  \\
        &\qquad \qquad -a_p t cos(W_p^{\max}t)+ a_e t cos(W_e^{\max}t) -\delta_2(\gamma) ] \\
        &= (V_e \!-\! a_et)sin\alpha sin(W_e^{\max}t) \!-\! (V_p \!-\! a_pt)sin\alpha sin(W_p^{\max}t)   \\
        &\quad \!+\!(V_p \!-\! a_pt)cos\alpha cos(W_p^{\max}t) \!-\! (V_e\!- \!a_et)cos\alpha cos(W_e^{\max}t) \\
        &\quad +sin \alpha\delta_1(\gamma) -cos\alpha\delta_2(\gamma) \\
        &= (V_e \!-\! a_et)cos(\alpha \!-\! W_e^{\max}t)\!-\!(V_p\!-\!a_pt)cos(\alpha-W_p^{\max}t) \\
        &\quad +sin \alpha\delta_1(\gamma) -cos\alpha\delta_2(\gamma).
    \end{aligned}
\end{equation*}
It is obvious that $\alpha-W_e^{\max}t<\alpha-W_p^{\max}t$ because $W_e^{\max}>W_p^{\max}$, and also $\dot {\Delta d_{pe}}<0$ if $\gamma$
is small enough. Without loss of generality, assume $V_e-a_et<V_p-a_pt$ as $t\in [0,\frac{\pi}{2W_p^{\max}}]$ by choosing $a_p$ and $a_e$ small enough. At $t=\frac{\pi}{2W_p^{\max}}$, $\alpha-W_p^{\max}t=\alpha-\pi/2\ge -\pi/2$, and $\alpha-3\pi/2 \le \alpha-W_e^{\max}t<\alpha-W_p^{\max}t$. For small enough $\gamma$, if $-3\pi/2 < \alpha-W_e^{\max}t<-\pi/2$, we have $cos(\alpha-W_e^{\max}t)< 0$, so $\dot {\Delta d_{pe}}<0$ because $cos(\alpha-W_p^{\max}t)\ge  0$. If $-\pi/2\le \alpha-W_e^{\max}t<\alpha-W_p^{\max}t$, we also have $\dot {\Delta d_{pe}}<0$ because $cos(\alpha-W_e^{\max}t)<cos(\alpha-W_p^{\max}t)$. Therefore, $\dot {\Delta d_{pe}}<0$ as $t\in [0,\frac{\pi}{2W_p^{\max}}]$ if $\gamma$ is small enough.

Integrating the derivatives from 0 to $t$ ($t\in [0,\frac{\pi}{2W_p^{\max}}]$) gives us the position

\begin{equation*}
    \begin{aligned}
        x_p(t)=&\frac{V_p}{W_p^{\max}}\big(cos(W_p^{\max}t)-1\big) - \frac{a_p}{W_p^{\max}} t cos(W_p^{\max}t) \\
        &+\frac{a_p}{(W_p^{\max})^2}sin(W_p^{\max}t) +\delta_3(\gamma), \\
        y_p(t)=& V_psin(W_p^{\max}t) -  \frac{a_p}{W_p^{\max}} t sin(W_p^{\max}t) +\frac{a_p}{(W_p^{\max})^2} \\
        &- \frac{a_p}{(W_p^{\max})^2}cos(W_p^{\max}t) +\delta_4(\gamma), \\
         x_e(t) =& \frac{V_e}{W_e^{\max}}\big(cos(W_e^{\max}t)-1\big) - \frac{a_e}{W_e^{\max}} t cos(W_e^{\max}t) \\
        &+\frac{a_e}{(W_e^{\max})^2}sin(W_e^{\max}t), \\
        y_e(t)=& d_0 +V_esin(W_e^{\max}t)  - \frac{a_e}{(W_e^{\max})^2}cos(W_e^{\max}t)\\
        &-  \frac{a_e}{W_e^{\max}} t sin(W_e^{\max}t) +\frac{a_e}{(W_e^{\max})^2},
    \end{aligned}
\end{equation*}
where $\delta_3(\gamma)\to 0$ and $\delta_4(\gamma)\to 0$ as $\gamma \to 0$.

At $T=\frac{\pi}{2W_p^{\max}}$, $x_p=-\frac{V_p}{W_p^{\max}} + \frac{a_p}{(W_p^{\max})^2}+\delta_3(\gamma)$, $y_p=V_p+(1-\frac{\pi}{2})\frac{a_p}{(W_p^{\max})^2}+\delta_4(\gamma)$. Since $W_p^{\max}<W_e^{\max}\le 3W_p^{\max}$ and $d_0\le \varepsilon_1$, it follows that
$x_e>-2\frac{V_e}{W_e^{\max}} - \frac{a_e}{(W_e^{\max})^2} =\bar x_e$ and $y_e<\varepsilon_1+V_e+2\frac{a_e}{(W_e^{\max})^2}+\frac{a_e\pi}{2W_p^{\max}W_e^{\max}}=\bar y_e$. 
We thus have $\Delta d_{pe}(\frac{\pi}{2W_p^{\max}}) < \sqrt{(x_p-\bar x_e)^2+(\bar y_e-y_p)^2}$ where 
$x_p-\bar x_e=-\frac{V_p}{W_p^{\max}} + \frac{a_p}{(W_p^{\max})^2}+2\frac{V_e}{W_e^{\max}} - \frac{a_e}{(W_e^{\max})^2} +\delta_3(\gamma)$ 
and 
$\bar y_e-y_p=\varepsilon_1+V_e+2\frac{a_e}{(W_e^{\max})^2}+\frac{a_e\pi}{2W_p^{\max}W_e^{\max}}-V_p-(1-\frac{\pi}{2})\frac{a_p}{(W_p^{\max})^2}-\delta_4(\gamma)$. Obviously $\sqrt{(x_p-\bar x_e)^2+(\bar y_e-y_p)^2}\le \varepsilon_2$ implies that $\Delta d_{pe}(\frac{\pi}{2W_p^{\max}})<\varepsilon_2$, thus the evader is captured at time T or earlier.

\textbf{Proof of Theorem~\ref{them2}}. 
Let us consider the following Lyapunov function
    $J(q_{di})=q_{di}^{\rm T}q_{di}.$
Taking (\ref{qdi}) into the derivative of $J(q_{di})$ gives
\begin{equation*}
    \dot J(q_{di})=-2(1-\frac{2}{1+e^{\|q_{di}\|^2-(d_{ic}^{des})^2}})\|q_{di}\|^2.
\end{equation*}
Denote $\Omega'=\{q_{di}:q_{di}=0\}$ and $\Omega_e=\{q_{di}:\|q_{di}\|=d_{ic}^{des}\}$. 
It is easy to see that $\dot J(q_{di})=0$ if and only if $q_{di}\in \Omega'\cup \Omega_e$.

As $\|q_{di}(t_0)\|> d_{ic}^{des}$, we have $\dot J(q_{di})<0$, so $\|q_{di}(t_0)\|$ shrinks. As $0<\|q_{di}(t_0)\| < d_{ic}^{des}$, $\dot J(q_{di})>0$. Therefore, by the LaSalle’s Invariance Principle \cite{khalil2002control}, as $\|q_{di}(t_0)\|>0$, we have $ q_{di}\to \Omega_e$ as $t\to \infty$.











\end{document}